\documentclass[aps,superscriptaddress,amsmath,amssymb,floatfix,twocolumn,showpacs,amsfonts,longbibliography,prb]{revtex4-1}
\usepackage{times}
\usepackage[varg]{txfonts}
\usepackage{textcomp}
\usepackage{graphicx}
\usepackage{subfigure}
\usepackage{tabu}
\usepackage{color}
\usepackage[plainpages=false,pdfpagelabels,colorlinks=true,linkcolor=red,urlcolor=blue,citecolor=blue,pdftitle={Title},pdfauthor={},pdfdisplaydoctitle=true,pdfduplex=DuplexFlipLongEdge]{hyperref}
\usepackage{braket}
\usepackage{float}
\usepackage{overpic}
\usepackage{gensymb}
\usepackage{mathrsfs}
\usepackage{tikz}
\usepackage{bm}
\usepackage{multirow}
\usepackage{amsfonts}
\usepackage{amsmath}
\usepackage{mathrsfs}
\usepackage{amssymb}
\usepackage{paralist}
\usepackage{indentfirst}
\usepackage{ulem}
\usepackage{soul}
\usepackage{cancel}
\usepackage{CJK}

\newcommand{\ketbra}[2]{\mathinner{|{#1}\rangle \langle{#2}|}}

\allowdisplaybreaks

\hypersetup{
           breaklinks=true,   % splits links across lines
           colorlinks=true,   % displays links as colored text instead of blocks
           pdfusetitle=true,  % \title and \author values into pdf metadata
        }
        
\newcommand{\tcr}{}

\begin{document}
\title{Quantum colored strings in the hole-doped $t$-$J_z$ model}
\author{Jia-Long Wang}
\affiliation{Beijing Computational Science Research Center, Beijing 100193, People's Republic of China}
\author{Shi-Jie Hu}
\email{Corresponding author: shijiehu@csrc.ac.cn}
\affiliation{Beijing Computational Science Research Center, Beijing 100193, People's Republic of China}
\affiliation{Department of Physics, Beijing Normal University, Beijing 100875, People's Republic of China}
\author{Xue-Feng Zhang}
\email{Corresponding author: zhangxf@cqu.edu.cn}
\affiliation{Department of Physics and Chongqing Key Laboratory for Strongly Coupled Physics, Chongqing University, Chongqing 401331, China}
\affiliation{Center of Quantum Materials and Devices, Chongqing University, Chongqing 401331, China}

\begin{abstract}
The stripe phase, an intertwined order observed in high-temperature superconductors, is regarded as playing a key role in elucidating the underlying mechanism of superconductivity, especially in cuprates.
Following Jan Zaanen's early scenario, %the fully-filled charge stripe
\tcr{the filled charge stripe, with one hole per unit cell of the charge order,}
can be taken as the interactive elastic quantum strings of holes, stabilized by $\pi$-phase shifts between neighboring magnetic domains.
However, this scenario is challenging to explain, particularly in terms of electron pairing, which necessitates hole pairs. In this work, we propose a new effective model for describing the stripe phase in the hole-doped $t$-$J_z$ model. With respect to the antiferromagnetic background, the model comprises three types of color-labeled point-defects coupling to an effective spin field, so named as {``colored string"}.
Comparing with numerical results from large-scale density matrix renormalization group (DMRG) simulations, we find semi-quantitative agreement in local hole density, magnetic moment, and the newly proposed spectrum features of the {effective} spin field. By systematically analyzing the hole-density distribution and the scaling of groundstate energy at different system sizes, we determine the effective core radius and the effective hopping amplitude of the quantum string. Furthermore, the local pinning field can be finely adjusted to drag the quantum string, offering a potential method for detecting it in optical lattices. At last, we further demonstrate the partially-filled stripe {with less than one hole per unit cell of the charge order} can also be well described by the effective theory. 
\end{abstract}
\maketitle

\section{Introduction}
%Stripe phase and one-dimensional defect
Decoding the mechanism of the high-$T_c$ superconductivity is one of the central tasks of condensed matter physics, also posing a challenge for the whole field of physics~\cite{hightc_review1, hightc_review2}.
Recent numerous numerical simulations of the Hubbard model~\cite{Hubbard_Science, Hubbard_PRL, Hubbard_PRX}, the $t$-$J$ model~\cite{t-J1, tj_jiang}, and their extensions~\cite{t-J_ext1, t-J_ext2, t-J_ext3, ttj_science, shiwei} have underscored the potential significance of the stripe phase~\cite{stripe_review1, stripe_review2, shiwei}.
This phase, observed prominently in the underdoped region, arises from the doping holes into the Mott insulator that hosts antiferromagnetic (AF) order~\cite{Exp_stripe1, Exp_stripe2, White_stripe, Zaanen_stripe}.
Then, the introduction of $\pi$-phase shifts between magnetic domains can lead to the stable emergence of one-dimensional ($1$D) topological defects, or strings~\cite{Zaanen01, string_Nishiyama}. 

%quantum string and 1-form charge
Strings widely exist in the frustrated magnetic system~\cite{string_peter}.
Because of the off-diagonal interactions, the string appears to behave in a ``quantum" manner, which means that the string becomes a worldsheet in the $2$+$1$ dimensions~\cite{Zaanen_stripe}.
The complicated interplay among these strings renders extremely unconventional quantum phases, including the deconfined quantum phase transition~\cite{dqcp01, dqcp02, dqcp03}, quantum string Luttinger liquid~\cite{zhou01, zhou02}, emergent RK point~\cite{changle}, and others~\cite{wanyuan, supersolid}.
These phenomena can be effectively described by a spin-$1/2$ chain within the framework of the effective theory~\cite{string_peter, dqcp01, supersolid, zhou02}.
Meanwhile, the quantum string can also be understood \textit{via} generalized symmetry~\cite{GS} and its relationship to the $1$-form charge with $d$-$1$ co-dimensions~\cite{string_xiong, mf_1form}.
Then, it would be straightforward to consider the possibility of an effective description of the string in the stripe phase. We can even believe that the quantum string could be the building block of cuprate high-$T_c$ superconductivity theory.

%Why t-Jz model
%The stripe phase can stably exist in the Hubbard model, t-J model, and their extensions.
%The best strategy to build an effective theory is to select the most simple model that hosts the stripe phase while avoiding some disturbing fluctuation that can not change the main physics.
%The Hilbert space is truncated from the Hubbard model to the $t$-$J$ model.
%Considering the intrinsic factor of the stripe phase is the $\pi$-phase shift between antiferromagnetic domains, we can further eliminate the quantum fluctuation of spin.
%Therefore, the best choice should be the $t$-$J_z$ model.

%Summary
{In this work, we propose an effective theory to describe the low-energy behaviors of hole-doped electronic systems, based on bottom-up modeling of the quantum string, referred to as ``colored string" (CS).}
%bottom-up construct {an effective model} capable of describing the quantum string{, referred to as ``colored string" (CS)}.%
%{In this work, we propose a theory to bottom-up construct an effective model 
% the model for the quantum string, referred to as ``colored string" (CS).}
To validate it, we consider the $t$-$J_z$ model, which contains the minimal ingredients required for the stripe phase to emerge as its ground state.
By employing the two-dimensional ($2$D) density matrix renormalization group (DMRG) method, we make a systematic comparison of numerical results.
At last, the effect of the pinning field and partially-filled string are discussed.
The paper is structured as follows: Sec.~\ref{sec2} introduces the model and quantum string, Sec.~\ref{sec3} introduces the effective model, Sec.~\ref{sec4} compares the DMRG results with those from the effective model, and finally, Sec.~\ref{sec5} presents our conclusion and discussion.

\section{Quantum String and model}\label{sec2}
\begin{figure}[t!]
\centering
\includegraphics[width=0.99\linewidth]{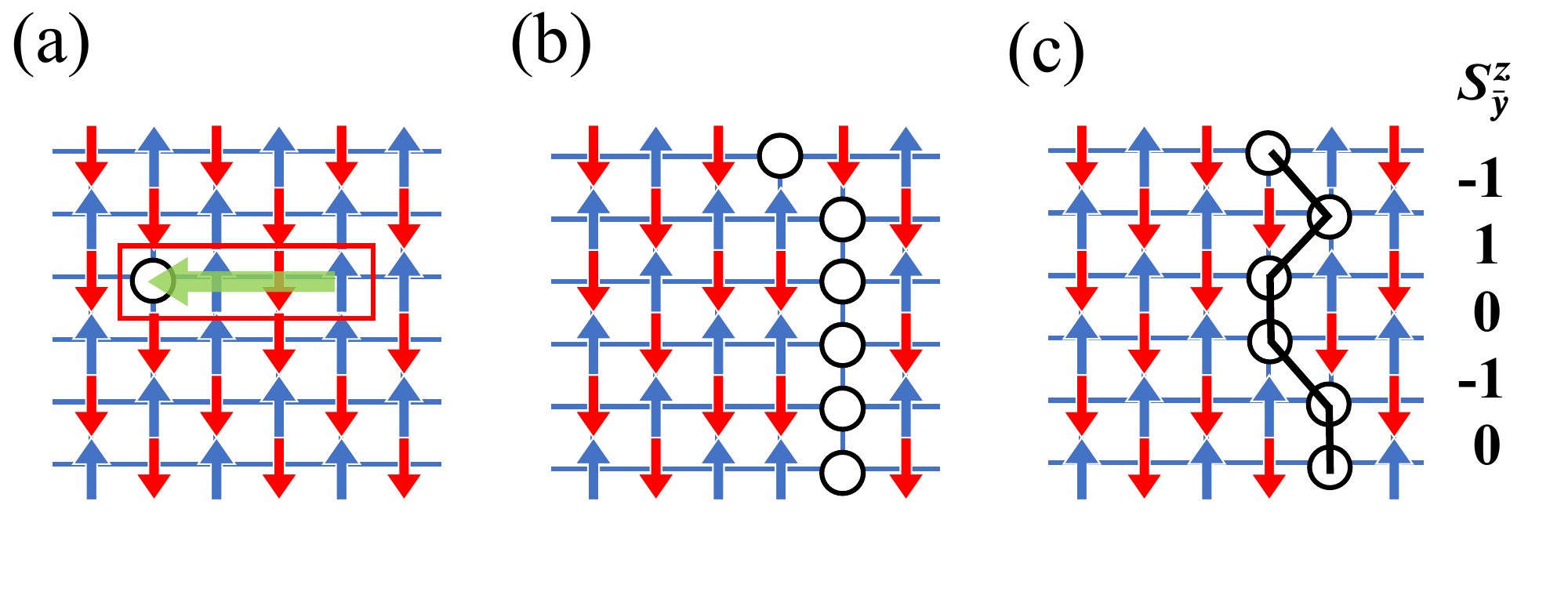}
\caption{Schematic pictures of (a) the Nagaoka mechanism, (b) a configuration of {$0$}-phase string, and (c) Zaanen's $\pi$-phase string (black line) {along with} its spin-$1$ representation.
The hole (black circle) moves in the AF background consisting of spin-up (blue arrow) and down (red arrow).
}\label{fig01}
\end{figure}

%Nagaoka mechanism
The ground state of cuprate without doping is the Mott insulator with antiferromagnetic (AF) spin order. When a single hole is doped into the AF background, it can move to another site through the hopping process. However, the creation of ferromagnetic bonds coincides with the breaking of AF bonds along the path of the hole, resulting in an energy cost proportional to the path length {[Fig.~\ref{fig01}(a)]}. This confinement of the hole by the AF background is usually known as the Nagaoka mechanism~\cite{nagaoka01, nagaoka02}.

%Quantum String
However, recent studies on the Hubbard model and the $t$-$J$ model~\cite{Hubbard_Science, Hubbard_PRL, Hubbard_PRX, t-J1, t-J_ext1, t-J_ext2, t-J_ext3} have shown that holes can induce the stripe phase, characterized by the stable presence of a $\pi$-phase shift between two neighboring magnetic domains.
According to Jan {Zaanen}'s scenario, these holes collectively form an elastic quantum string.
Subject to the topological constraint, two domains exhibit either a $0$ or $\pi$-phase shift, with the string identified as a $0$-phase or $\pi$-phase string, respectively.
In comparison, the $\pi$-phase string can vibrate more readily through hole hopping and has lower energy compared to the $0$-phase string, so that the $\pi$-phase string usually appears to be more stable in cuprate related models {[Figs.~\ref{fig01}(b,c)]}.
Most importantly, the incommensurate magnetic order observed in experiments can be attributed to the effective repulsive interaction between strings~\cite{Zaanen_stripe}.
{In this work, we refer to each hole stripe as a CS, emphasizing the fundamental role of the CS in our effective theory.
For instance, we use the term ``filled CS" rather than ``filled stripe".}

%Introduction of the model
The simplest Hamiltonian which hosts the physics above is the $t$-$J_z$ model on a square lattice written as follows:
\begin{equation}
H=-t \sum_{\langle i,j\rangle} \sum_\sigma \left(f_{i, \sigma}^\dag f^{\phantom{\dag}}_{j, \sigma} + \textrm{h.c.}\right) - \frac{J_z}{2} \sum_{\langle i, j\rangle} \sum_\sigma n^{\phantom{\dag}}_{i, \sigma} n^{\phantom{\dag}}_{j, \bar{\sigma}}\ ,
\label{eq1}
\end{equation}
where $f_{i, \sigma}^\dag$, $f^{\phantom{\dag}}_{i, \sigma}$ and $n^{\phantom{\dag}}_{i, \sigma} = f^\dag_{i, \sigma} f^{\phantom{\dag}}_{i, \sigma}$ represent the creation, annihilation, and density operator of the fermion with the spin polarization $\sigma=\uparrow$, $\downarrow$ along the $z$-axis, respectively.
{It is important to note that double-occupancies are prohibited at all sites.
The basis set for site $i$ consists of $\ket{\text{vac}_i}$, $\ket{\uparrow_i} = f_{i, \uparrow}^\dag \ket{\text{vac}_i}$, and $\ket{\downarrow_i} = f_{i, \downarrow}^\dag \ket{\text{vac}_i}$.}
$\bar{\sigma}$ is the reverse of $\sigma$.
$t$ denotes the coefficient of the hoppings between two NN sites labeled as the bond $\langle i,j\rangle$, and $J_z$ give the magnitude of the Ising-type interactions.
Hereafter, $t=1$ sets the energy unit.
$L_x$ and $L_y$ represent the length and the circumference of the cylinder, respectively.
The coordinate of site-$i$ is given by $(x_i,\ y_i)$, where $x_i=1$, $\cdots$, $L_x$ and $y_i = 1$, $\cdots$, $L_y$.
If necessary, we will change the index labels of the sites in the following discussion, e.g., $n_{i,\sigma} \equiv n_\sigma (x_i, y_i)$.
The boundary condition of the cylinders is periodic (PBC) in the $y$-axis while open (OBC) in the $x$-axis.
The number of lattice sites $N=L_x L_y$ is also defined. Here, we only study a single string, so the filling number of holes is chosen as $N^{(\text{h})} \le L_y$ with respect to the AF background. In the following section, we will review Zaanen's string as a primer before discussing the novel colored string.

%\subsection{Zaanen's string}
%short review
When $J_z$ dominates, $L_y$ holes doped into the AF background tend to stay together to minimize potential energy dictated by the $J_z$-terms in the model~\eqref{eq1}. they form a block cluster with the largest number of adhering AF bonds.
Conversely, when $t$ is significant, collective hole motion is absent {as the system prefers the metallic phase}.
In the intermediate region, a {$\pi$-phase} string can survive in the ground state, thereby separating the whole system into two AF domains~\cite{Zaanen01}.
{According to the energy comparison presented in Appendix~\ref{AppA1}, the ground state for $J_z / t \le 3.8$ is precisely situated in the intermediate region.}

{To align with the empirically estimated ratio between the Coulomb repulsion $U$ and the hopping coefficient $t$ for typical cuprates, which is $U/t=8\sim12$~\cite{hubbard_U_ab}, the standard $t$-$J$ model typically has $J_z/t \in [1/3,\, 0.5]$.
In the $t$-$J_z$ model, the energy gap between the AF and ferromagnetic bonds is $J_z/2$, whereas in the $t$-$J$ model, it is $J$.
So the ratio $J_z / t$ for the $t$-$J_z$ model is approximately twice that of $J/t$ for the $t$-$J$ model, i.e., $J_z/t \in [2/3,\, 1]$.
In the following discussion, we focus on two typical values of the ratio $J_z / t = 0.6$ and $1$.}

As with the string in frustrated magnetism~\cite{string_peter, dqcp01, supersolid,zhou02}, the dynamics of the string can be effectively described by the spin-$1$ chain under the assumption that row-$y$ has a single hole located at $\mathcal{X}^{(\text{h})}_y$.
Specifically, we define a spin operator $\mathbf{S}_{\bar{y}}$ between neighboring row-$y$ and row-$(y + 1)$, comprising components $S^+_{\bar{y}}$, $S^-_{\bar{y}}$ and $S^z_{\bar{y}}$, which represent the {standard} spin-flipping-up and down operators as well as the $z$-axis polarization {for spin-$1$}, respectively.
Obviously, $\bar{y}=y+1/2$ modulo $L_y$.
When the $x$-coordinates of the holes in these two rows are the same, that is, $\mathcal{X}^{(\text{h})}_y = \mathcal{X}^{(\text{h})}_{y+1}$, the configuration corresponds to ${\tikz[baseline={(0,-0.05)}]{\draw (0,-0.15) -- (0,0.15);\draw (0,-0.15) circle (1.mm);\draw (0,0.15) circle (1.mm);}}$ and is represented by a non-magnetic state $\ket{S^z_{\bar{y}} = 0} \equiv \ket{0_{\bar{y}}}$ {[Fig.~\ref{fig01}(c)]}.
Therefore, the straight string along the $y$-axis corresponds to the state $\ket{0_{1/2} 0_{3/2} \cdots 0_{L_y-1/2}}$.
Similarly, we can map the configuration ${\tikz[baseline={(0,-0.05)}]{\draw (0,0.15) circle (1.mm);\draw (0.2,-0.15) circle (1.mm);\draw (0,0.15) -- (0.2,-0.15);}}$ to $\ket{1_{\bar{y}}}$, and ${\tikz[baseline={(0,-0.05)}]{\draw (0,-0.15) circle (1.mm);\draw (0.2,0.15) circle (1.mm);\draw (0,-0.15) -- (0.2,0.15);}}$ to $\ket{(-1)_{\bar{y}}}$.
In {Fig.~\ref{fig01}(c)}, the hole hopping processes mirror the effective spin-exchanging terms $\left(S^+_{\bar{y}} S^-_{\bar{y}+1} + \textrm{h.c.}\right)$.
Additionally, both the configurations $\ket{\pm 1}$ violate an AF bond along the $y$-axis, resulting in a single-ion anisotropic term $\left(S_{\bar{y}}^z\right)^2$.
%It can be reasonably assumed that ferromagnetic bonds along the $x$-axis are absent if the displacement of the holes in two NN rows along the $x$-axis $\mathcal{D}^{(\text{h})}_{\bar{y}} = \mathcal{X}^{(h)}_{y+1} - \mathcal{X}^{(h)}_y$ is less than or equal to $1$.

Overall, the effective spin-$1$ Hamiltonian is given by 
\begin{equation}
\nonumber
    H_\text{e} = -\frac{t}{2} \sum_y \left(S_{\bar{y}}^+ S_{\bar{y}+1}^- + \textrm{h.c.}\right) + \frac{J_z}{2} \sum_y \left(S_{\bar{y}}^z\right)^2.
    %\label{eq2}
\end{equation}
where the ratio between $J_z$ and $t$ determines the ground state.
Notably, the summation over $y$ is equivalent to that over $\bar{y}$, with the former being the preferred choice in this study.
%In fact, the model~\eqref{eq2} describes a $\pi$-phase string with two fixed endpoints.
%{To assume that an endpoint is located at the first row with the $x$-coordinate of $\mathcal{X}^{(\text{h})}_1$, it is straightforward to demonstrate that the other fixed endpoint of the string is located at row-$L_y$ and its $x$-coordinate is given by $\mathcal{X}^{(\text{h})}_{L_y} = \mathcal{X}^{(\text{h})}_1 - \sum_y S^z_{\bar{y}}$.}
Similar to the Haldane phase~\cite{Chen_HI}, $\ket{\pm 1}$ mark two point-like topological defects that cannot be annihilated \textit{via} local operation.
They serve as the $0$-form topological charge on the $1$-form string~\cite{ice_sac}.
However, Zaanen's string is so simple that some interesting processes are lost, e.g., the electron pairing.

\section{Effective colored string theory}\label{sec3}

\begin{figure}[t]
\centering
\includegraphics[width=0.99\linewidth]{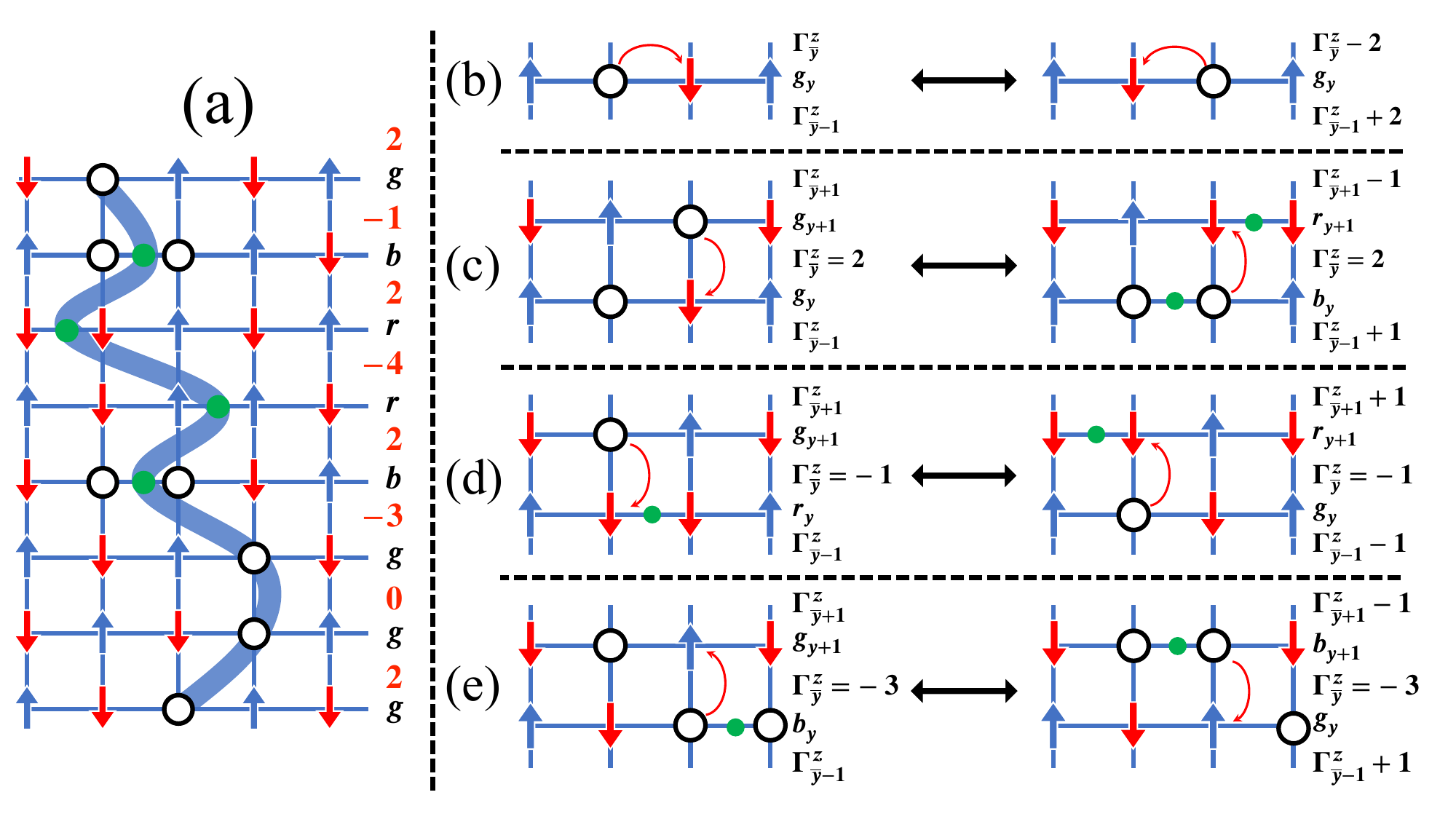}
\caption{(a) colored particles carry a effective spin fields. (b) illustration of off-diagonal interactions $H_\text{o}^{(\textbf{g})}$. (c) illustration of $H^{(\textbf{r}\text{-}\textbf{b} / \textbf{g}\text{-}\textbf{g})}_\text{o}$. (d) illustration of $H^{(\textbf{r}\text{-}\textbf{g}/\textbf{g}\text{-}\textbf{r})}_\text{o}$. (e) illustration of $H^{(\textbf{b}\text{-}\textbf{g}/\textbf{g}\text{-}\textbf{b})}_\text{o}$.
 }\label{fig022}
\end{figure}

%\subsection{Colored string}
%Matter filed and on-site interaction
The fundamental unit of Zaanen's string is the $\pi$-phase holon ($\uparrow\!\!\circ\!\!\downarrow$ and $\downarrow\!\!\circ\!\!\uparrow$).
However, with the allowance of the hole hoppings along the $y$-axis, both the spinon ($\uparrow\downarrow\downarrow\uparrow$ and $\downarrow\uparrow\uparrow\downarrow$) and the dual-hole ($\uparrow\!\!\circ\circ\!\!\uparrow$ and $\downarrow\!\!\circ\circ\!\!\downarrow$) emerge {in CS [Fig.~\ref{fig022}(a)]}.
For simplicity, we use the colors $c = \textbf{r}$, $\textbf{g}$, and $\textbf{b}$ to label the spinon, holon, and dual-hole, respectively, and refer to them as \textit{color particles} (CP).
Thus, the annihilation (creation) operators for these particles at row-$y$ are denoted as $\text{r}^{\phantom{\dag}}_y$, $\text{g}^{\phantom{\dag}}_y$, and $\text{b}^{\phantom{\dag}}_y$ ($\text{r}^\dag_y$, $\text{g}^\dag_y$, and $\text{b}^\dag_y$), using the mapping of \textit{hardcore bosons}.
Their density operators are given by $n^{(\textbf{r})}_y = \text{r}^\dag_y \text{r}^{\phantom{\dag}}_y$, $n^{(\textbf{g})}_y = \text{g}^\dag_y \text{g}^{\phantom{\dag}}_y$ and $n^{(\textbf{b})}_y = \text{b}^\dag_y \text{b}^{\phantom{\dag}}_y$.
{To create a CP, it is necessary to violate a certain amount of AF bonds in the $x$-axis, referred to as the \textit{onsite energy} [Table~\ref{tabel1}].}

{It is noted that the interpretation of three CPs differs somewhat from those in the previous studies.
For example, conventional ``spinons" are typically associated with fractional statistics within the parton construction, whereas we define them here as hardcore bosons.
And then we put the signs related to the exchange of electrons during hopping processes directly into the effective hopping coefficients.}

Moreover, we impose a constraint permitting only one CP in row-$y$: $\sum_{c} n^c_y = 1$, with the CP positioned at $x=\mathcal{X}^{(\text{CP})}_y$.
{Thus, the basis set for row-$y$ consist of $\ket{\textbf{r}}$, $\ket{\textbf{g}}$, and $\ket{\textbf{b}}$.
Within this framework, we can define the particle number operators for the color particles as follows
\begin{equation}\label{eq:colorparticlenumberops}
\begin{split} 
n^{(\textbf{r})}_y
=\begin{pmatrix} 
1 & 0 & 0 \\
0 & 0 & 0 \\
0 & 0 & 0 
\end{pmatrix}\, ,\quad
n^{(\textbf{g})}_y
=\begin{pmatrix} 
0 & 0 & 0 \\
0 & 1 & 0 \\
0 & 0 & 0 
\end{pmatrix}\, ,\quad
n^{(\textbf{b})}_y
=\begin{pmatrix} 
0 & 0 & 0 \\
0 & 0 & 0 \\
0 & 0 & 1 
\end{pmatrix}\, .
\end{split}
\end{equation}
Additionally, we define the color-exchanging operators, for example,
\begin{equation}\label{eq:colorexchangeops}
\begin{split} 
\text{r}^\dag_y \text{g}^{\phantom{\dag}}_y
=\begin{pmatrix} 
0 & 1 & 0 \\
0 & 0 & 0 \\
0 & 0 & 0 
\end{pmatrix}\, ,\,
\text{r}^\dag_y \text{b}^{\phantom{\dag}}_y
=\begin{pmatrix} 
0 & 0 & 1 \\
0 & 0 & 0 \\
0 & 0 & 0 
\end{pmatrix}\, ,\,
\text{g}^\dag_y \text{b}^{\phantom{\dag}}_y
=\begin{pmatrix} 
0 & 0 & 0 \\
0 & 0 & 1 \\
0 & 0 & 0 
\end{pmatrix}\, ,
\end{split}
\end{equation}
Notably, all terms of the new effective Hamiltonian $H_\text{e}^\text{CS}$ always appear in pairs of hardcore bosonic operators of CPs for row-$y$, and thus the exchange between operators of two rows does not introduce a negative sign, reflecting the hardcore bosonic nature of these particles.

As the midpoint of the spinon \textbf{r} and the dual-hole \textbf{b} sit at the bond, we set the lattice spacing to $2$ in this work, ensuring that the $x$-coordinates of CPs are integer [Fig.~\ref{fig022}].
The displacement along the $x$-axis between the holons at row-$y$ and row-$(y+1)$, i.e., $\mathcal{X}^{(\text{CP})}_{y+1} - \mathcal{X}^{(\text{CP})}_y$, can be represented as {an effective} \textit{spin field} $\Gamma^z_{\bar{y}} \in \mathbb{Z}$ with integer values, where its absolute value satisfies $\vert \Gamma^z_{\bar{y}} \vert \le \bar{\Gamma}$, with $\bar{\Gamma} \ge 1$ defining the upper bound of $\vert \Gamma^z_{\bar{y}} \vert$.
We write down a basis set of $\{\ket{\Gamma^z_{\bar{y}}}\}=\{\ket{\bar{\Gamma}}$, $\cdots$, $\ket{-\bar{\Gamma}}\}$, and the operators for the spin field can be explicitly written as
\begin{equation}\label{eq:colorexchangeops}
\begin{split}
\Gamma^z_{\bar{y}}
\!=\!\begin{pmatrix}
\bar{\Gamma} & & \\
 & \ddots & \\
 & & -\bar{\Gamma} \\
\end{pmatrix} ,\ 
\Gamma^+_{\bar{y}}
 \!=\!\begin{pmatrix}
 & 1 & & \\
 & & \ddots & \\
 & & & 1 \\
 & & & \\
\end{pmatrix} ,\ 
\Gamma^-_{\bar{y}}
\!=\!\begin{pmatrix}
 & & & \\
1 & & & \\
 & \ddots & & \\
 & & 1 & \\
\end{pmatrix} .\
\end{split}
\end{equation}
In particular, the ``shift operator" $\Gamma^+_{\bar{y}}$ ($\Gamma^-_{\bar{y}}$) which raises (lowers) spin field as $\Gamma^\pm_{\bar{y}} \ket{\Gamma^z_{\bar{y}}} = \ket{\Gamma^z_{\bar{y}} \pm1}$.

In the effective theory, we use a convention for the ordering of lattice sites: the lattice sites are arranged in one-dimensional ($1$D) path, with the coordinates $(x,\ y) = (L_x,\ 1)$, $\cdots$, $(1,\ 1)$, $(L_x,\ 2)$, $\cdots$, $(1,\ 2)$, $\cdots$.
As a result, the Hamiltonian of the colored string in $t$-$J_z$ model consists of two categories:
\begin{equation}
\begin{split}
H^\text{CS}_\text{e}\! &= \! H_\text{d} \! + \! \left(H_\text{o}^{(\textbf{g})} \! + \! H^{(\textbf{r}\text{-}\textbf{b} / \textbf{g}\text{-}\textbf{g})}_\text{o} \! + \! H^{(\textbf{r}\text{-}\textbf{g}/\textbf{g}\text{-}\textbf{r})}_\text{o} \! + \! H^{(\textbf{b}\text{-}\textbf{g}/\textbf{g}\text{-}\textbf{b})}_\text{o} \! + \! \textrm{h.c.}\right)
\label{effHam}
\end{split}\, .
\end{equation}
$H_{\text{d}}$ represents the \textit{diagonal interactions}, which are influenced by the distance between CPs, and latter represent \textit{off-diagonal interactions}, which deform the CS, produce the \textbf{r}-\textbf{b} pair, and move CPs along the CS.

\textit{Diagonal interactions}---The violation of the AF bonds along the $y$-axis results in the diagonal interaction among CPs.
After checking the violated AF bonds, we clearly find that the energy loss is proportional to the distance $\vert \Gamma^z_{\bar{y}} \vert$.
This means that configurations with larger distances have higher energy values and are thus significantly discouraged in the ground state, reminiscent of the Nagaoka mechanism.
Then, the sign of $\Gamma^z_{\bar{y}}$ depends on the relative direction between CPs, and it does not affect the interaction.
The same features can also be found in the interactions between different species of CPs [Appendix~\ref{App:A2}].

After calculating all interactions [Table~\ref{tabel2}], we can derive the diagonal energy of a CS, i.e.,
\begin{eqnarray}
H_\text{d} = \sum_y \sum_{c, c'} V_{c,c'} \left( \lvert \Gamma^z_{\bar{y}} \rvert \right) n^{(c)}_y n^{(c')}_{y+1}\ .\nonumber
\label{}
\end{eqnarray}

\textit{Off-diagonal interactions}---The shape of the CS changes through the hopping processes of holes, which contributes to the vibrational energy.
Same as Zaanen's string, when a hole in row-$y$ hops to the left or the right, the effective spin fields in neighboring dual rows $(\bar{y} - 1)$ and $\bar{y}$ change as well [Fig.~\ref{fig022}(b)].
We can derive the operator representation of the off-diagonal interactions [Figs.~\ref{fig022}(b-e)], given by
\begin{subequations}
\begin{align}
\nonumber
H_\text{o}^{(\textbf{g})} &= -t \sum_y \left[ \left(\Gamma^+_{\bar{y}}\right)^2 n^{(\textbf{g})}_y \left(\Gamma^-_{\bar{y}-1} \right)^2 + \textrm{h.c.}\right]\,  \\
\nonumber
H^{(\textbf{r}\text{-}\textbf{b} / \textbf{g}\text{-}\textbf{g})}_\text{o} &=\pm t \sum_y \left(\Gamma^\mp_{\bar{y}+1} \text{r}_{y+1}^\dag \text{b}_y^\dag \Gamma^\pm_{\bar{y}-1} + \Gamma^\pm_{\bar{y}+1} \text{b}_{y+1}^\dag \text{r}_y^\dag \Gamma^\mp_{\bar{y}-1}\right) \text{g}^{\phantom{\dag}}_{y+1} \text{g}^{\phantom{\dag}}_y\, \\
\nonumber
H^{(\textbf{r}\text{-}\textbf{g}/\textbf{g}\text{-}\textbf{r})}_\text{o} &= -t \sum_y \Gamma^\mp_{\bar{y}+1} \text{r}_{y+1}^\dag \text{r}^{\phantom{\dag}}_y \text{g}_y^\dag \text{g}^{\phantom{\dag}}_{y+1} \Gamma^\pm_{\bar{y}-1}\, ,
\hspace{10px} \Gamma_{\bar{y}}^z=\pm1\, \\
\nonumber
H^{(\textbf{b}\text{-}\textbf{g}/\textbf{g}\text{-}\textbf{b})}_\text{o} &= -t \sum_y \Gamma^\pm_{\bar{y}+1} \text{b}_{y+1}^\dag \text{b}^{\phantom{\dag}}_y \text{g}_y^\dag \text{g}^{\phantom{\dag}}_{y+1} \Gamma^\mp_{\bar{y}-1}\, , \hspace{7px} \Gamma_{\bar{y}}^z=\pm3\, 
\end{align}
\label{eqn04}
\end{subequations}
Then, we can establish a connection with Zaanen's string by noting that the spin field reduces to the one for the spin-$1$ through the relations $\left( \Gamma^\pm_{\bar{y}} \right)^2 \leftrightarrow S^\pm_{\bar{y}}/\sqrt{2}$ and $\Gamma^z_{\bar{y}} \leftrightarrow 2 S^z_{\bar{y}}$, where $\bar{\Gamma} = 1$.
{$H_\text{o}^{(\textbf{g})}$ represents the movement of \textbf{g} particle along the $x$-axis, caused by hole hopping processes along the $x$-axis. While $H^{(\textbf{r}\text{-}\textbf{b} / \textbf{g}\text{-}\textbf{g})}_\text{o}$, $H^{(\textbf{r}\text{-}\textbf{g}/\textbf{g}\text{-}\textbf{r})}_\text{o}$,
$H^{(\textbf{b}\text{-}\textbf{g}/\textbf{g}\text{-}\textbf{b})}_\text{o}$ generate \textbf{r}, \textbf{b} particles by hole hopping processes along $y$-axis.}

In addition, we define the translation of the CS along the $x$-axis [Appendix~\ref{App:A3}], which allows CS to reach the leftmost or rightmost edges.}

\section{Numerical Results}\label{sec4}

To analyze the groundstate properties of CS, we implement exact diagonalization (ED) on the effective model Hamiltonian $H^\text{CS}_\text{e}$.
The Hilbert space bases are fully labeled by the color set $\{c\}$ of CPs for all rows, the corresponding spin-field set $\{\Gamma^z\}$, and the location $\mathcal{X}^{(\text{CS})}$ of CS, i.e.,
\begin{equation}
\ket{\{c\}, \{\Gamma^z\}, \mathcal{X}^{(\text{CS})}} = \left(\bigotimes^{L_y}_{y=1} \ket{c_y}\right) \left(\bigotimes^{L_y}_{y=1} \ket{\Gamma^z_{\bar{y}}} \right) \ket{\mathcal{X}^{(\text{CS})}}\, .\nonumber
\end{equation}
For example, the string configuration depicted in Fig.~\ref{fig022}(a) is labeled as \{c\} = \{\textbf{g}, \textbf{g}, \textbf{g}, \textbf{r}, \textbf{b}, \textbf{b}, \textbf{r}, \textbf{g}\}, $\{\Gamma^z\} = \{2$, $0$, $-3$, $2$, $-4$, $2$, $-1$, $2\}$, and $\mathcal{X}^{(\text{CS})}=3$.

{For a fixed position of the colored string $\mathcal{X}^{(\text{CS})}$, we construct the Hilbert space by taking the direct product of the bases for color particles and spin fields.
Importantly, during this process, we must account for the gauge constraints.
For instance, for spinons (\textbf{r} particles) located at two adjacent rows $y$ and $y+1$, the spin field operator $\Gamma^z_{\bar{y}}$ can only assume even integer values.
In contrast, when one spinon is replaced by a holon, $\Gamma^z_{\bar{y}}$ is restricted to odd integer values.
By summing over all possible positions of colored string, we can establish a one-by-one correspondence between the bases used in our effective colored string theory and those employed in the fermionic $t$-$J_z$ model.}

The absolute values of spin fields are capped at $\vert \Gamma^z_{\bar{y}} \vert \le 8$.
and the equalities $\sum^{L_y}_{y=1} n^\textbf{r}_y = \sum^{L_y}_{y=1} n^\textbf{b}_y$ and $\sum_y \Gamma^z_{\bar{y}} = 0$ hold.
And thus, for $L_x=11$ and $L_y=6$, the size of the Hilbert space reaches up to $18,601,989$.
The wave function can be spanned as a superposition
\begin{equation}
\nonumber
\ket{\psi} = \sum_{\{c\}, \{\Gamma^z\}, \mathcal{X}^{(\text{CS})}} \psi \left(\{c\}, \{\Gamma^z\}, \mathcal{X}^{(\text{CS})}\right) \ket{\{c\}, \{\Gamma^z\}, \mathcal{X}^{(\text{CS})}}\, .
\end{equation}

%DMRG
The brutal force of $2$D DMRG method is also employed to investigate the CS motion in the $t$-$J_z$ model~\eqref{eq1} {, the hole concentration for the case of a single stripe is given by $\nu/L_x$, where $\nu \le 1$ represents the hole filling fraction within the stripe.}
Unless otherwise stated, we set the bond dimension $\chi=8,192$ to ensure truncation errors below $10^{-5}$ for a cylinder with $L_x \le 23$ and $L_y=6$ {[Appendix~\ref{AppB}]}.
We choose two distinct strengths for the Ising-type interaction $J_z = 0.6$ and $1$, where a $\pi$-phase string is favored in the ground state.
In addition, we adjust $J_z$ twice for bonds connected to sites at two edge columns to prevent the formation of a frozen string.
{We use another convention for the ordering of lattice sites in $2$D DMRG: the lattice sites are arranged in one-dimensional ($1$D) ``snake" path~\cite{2ddmrg}, with the coordinates $(x, y) = (1,\ 1)$, $\cdots$, $(1,\ L_y)$, $(2,\ 1)$, $\cdots$, $(2,\ L_y)$, $\cdots$.
When comparing results, we always transform the convention used in DMRG to align with that used in our effective theory.}

\begin{figure}[t]
\centering
\includegraphics[width=0.99\linewidth]{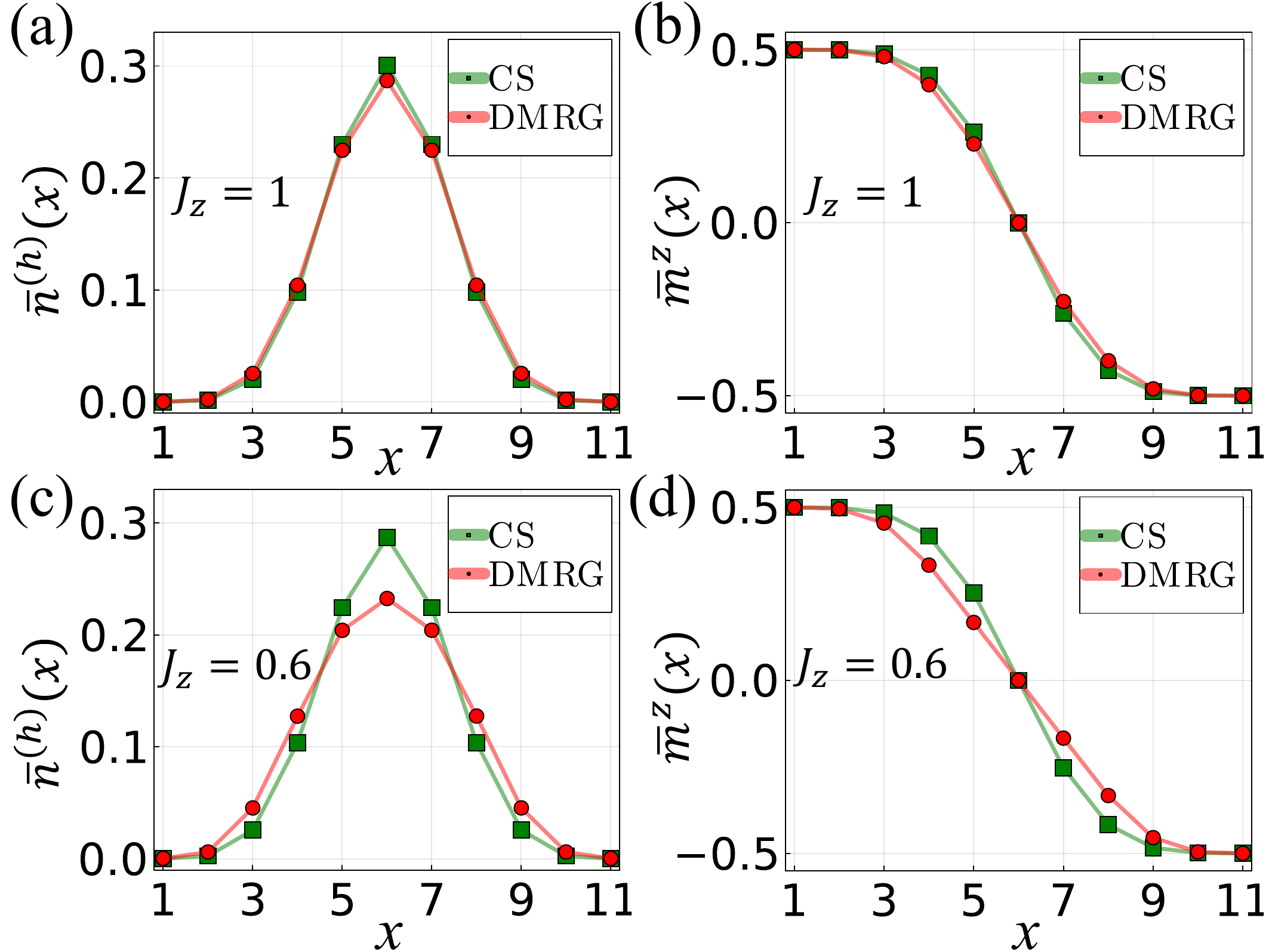}
\caption{{The distribution of (a,c) the average hole density $\bar{n}^{(h)}(x)$ and (b,d) average magnetic moment $\bar{m}^z(x)$ alone x-direction in a $11 \times 6$ cylinder of filled CS.
A comparison is made between the ED results of the colored string Eq.~\eqref{effHam} and the DMRG results of the $t$-$J_z$ model Eq.~\eqref{eq1} at $J_z=0.6$ and $J_z=1$.}
 }\label{fig05}
\end{figure}

\subsection{Hole-density distribution}
%density distribution
At first, we consider %the fully-filled CS
{a filled CS (i.e., $\nu=1$)}. 
For a cylinder with a finite length, the hole-density distribution provides a signal of CS~\cite{Hubbard_Science, tj_jiang}.
With defining the local hole density $n^{(\text{h})} (x,\ y) = \braket{1 - n_\uparrow (x,\ y) - n_\downarrow (x,\ y)}$ at site-$i=(x,\ y)$, we get the average hole density {[Figs.~\ref{fig05}(a,c)]}
\begin{equation}
\bar{n}^{(\text{h})} (x) = \frac{1}{L_y} \sum^{L_y}_{y=1} n^{(\text{h})} (x,\ y)\, ,\nonumber
\end{equation}
which indicates that holes tend to congregate at the cylinder center.
Similarly, with the local magnetic moment $m^z (x,\ y) = (1/2)\braket{n_\uparrow (x,\ y) - n_\downarrow (x,\ y)}$, we compute the average magnetic moment {[Figs.~\ref{fig05}(b,d)]}
\begin{equation}
\nonumber
\bar{m}^z (x) = \frac{1}{L_y} \sum^{L_y}_{y=1} (-1)^{x+y} m^z (x,\ y)
\end{equation}
showing a clear $\pi$-phase shift in the AF background.
At large $J_z=1$, the effective model demonstrates a semi-quantitative agreement with the DMRG numerical results, thereby providing strong support for the CS picture.
In comparison, the DMRG result at $J_z=0.6$ shows that the hole-density pack becomes slightly broader.

\subsection{Spectra of spin fields}
%Gauge field
In any given row-$\bar{y}$, the spin field $\Gamma^z_{\bar{y}}$ displays statistical features in the reduced density matrix
\begin{equation}
\nonumber
\begin{split}
\rho_{\bar{y}} &= \text{tr}_\text{p} \ketbra{\psi}{\psi}\\
&= \sum_{c_y, c_{y+1}, \Gamma^z_{\bar{y}}} \Lambda \left(c_y, c_{y+1}, \Gamma^z_{\bar{y}}\right)\ket{c_y c_{y+1}} \ketbra{\Gamma^z_{\bar{y}}}{\Gamma^z_{\bar{y}}} \bra{c_y c_{y+1}}
\end{split}
\end{equation}
after other degrees of freedom $\text{p}$ are traced out.
The resulting spectrum weight $\Lambda$ is referred to as the \textit{fingerprint} of CS.
We notice that the spectrum is independent on the choice of $\bar{y}$ owing to the translation symmetry along the $y$-axis, so we simplify the definition to $\Lambda \left(c, c', \Gamma^z \right) \equiv \Lambda \left(c_y, c_{y+1}, \Gamma^z_{\bar{y}}\right)$, where both $c$ and $c'$ represent color values.
%which has effectively ruled out the possibility of a ``random walk" picture of uncorrelated holes near the center of the cylinder.
%Definitely, people can argue that such phenomena can also be explained as uncorrelated holes randomly walking in this cylinder system. 
%Thus, the key feature would be the structure of the string, such as the distribution of the effective spin fields.
%Here, we focus on the effective spin fields between different types of CPs, because they reflect more intrinsic features of string. 
%At first, as shown in Fig.~\ref{fig06}(a,d),  the probability is strongly suppressed by the magnitude of the spin field $\vert \Gamma^z \vert$, which has been verified by DMRG calculations as well.
%which indicates CPs form a curve-like topological defect meaning string, 

\begin{figure}[t]
\centering
\includegraphics[width=0.99\linewidth]{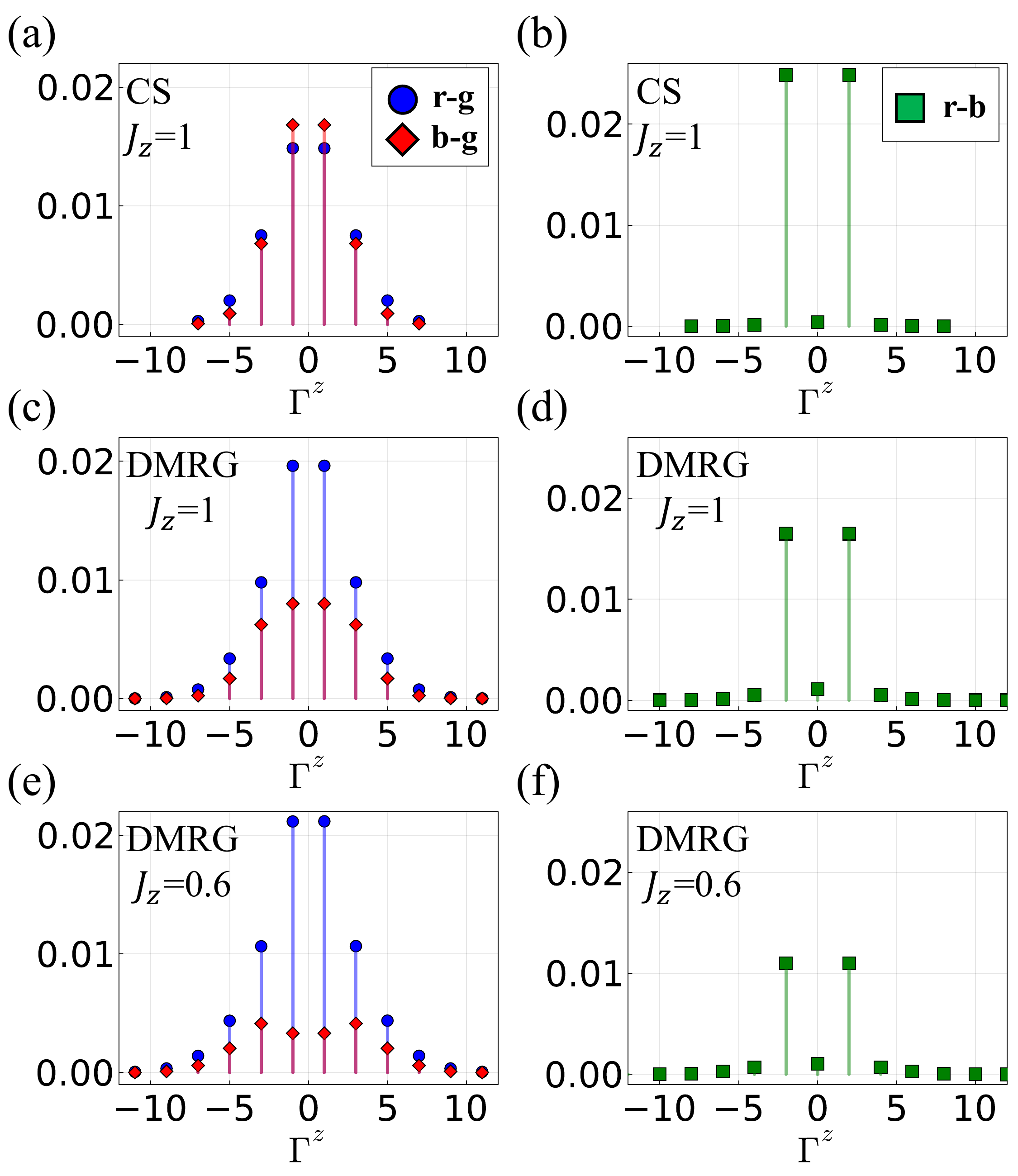}
\caption{The spectrum weight $\Lambda \left(c, c', \Gamma^z \right)$ as a function of the spin field $\Gamma^z$ in a $11 \times 6$ cylinder at (a-d) $J_z=1$ and (e,f) $0.6$. Both the channels (a,c,e) \textbf{r}-\textbf{g} and \textbf{b}-\textbf{g}, as well as (b,d,f) \textbf{r}-\textbf{b}, are investigated.
A comparison is made between (a,b) the ED results of the effective model~\eqref{effHam}, and (c-f) the DMRG results of the $t$-$J_z$ model~\eqref{eq1}.
 }\label{fig06}
\end{figure}

The statistical features are summarized below: (\textbf{i}) For channels \textbf{r}-\textbf{g} and \textbf{b}-\textbf{g}, spectrum weights primarily distribute at $\Gamma^z=\pm1$ and $\pm3$, while \textbf{r}-\textbf{b} shows concentration at $\Gamma^z$=$\pm2$ {[Figs.~\ref{fig06}(a-d)]}.
Remarkably, this phenomenon persists even at a lower value of $J_z=0.6$ {[Figs.~\ref{fig06}(e,f)]}.
The concentration of the spectrum weight at $\Gamma^z$=$\pm2$ for \textbf{r}-\textbf{b} suggests the confinement of spinon and dual-hole.
(\textbf{ii}) Additionally, deviations from the effective theory are noticeable in the spectra for \textbf{r}-\textbf{g} and \textbf{b}-\textbf{g} {[Figs.~\ref{fig06}(c,e)]}.
In comparison to \textbf{b}-\textbf{g}, the increased weight for \textbf{r}-\textbf{g} implies higher spinon mobility compared to dual-hole.
This is reasonable because the \textbf{r}-\textbf{g} channel uses the bases with $\Gamma^z=\pm1$ {[Figs.~\ref{fig04}(a,b)]}.
(\textbf{iii}) When considering the exchanging process between \textbf{b} and \textbf{g} in the off-diagonal interaction $H^{(\textbf{b}\text{-}\textbf{g}/\textbf{g}\text{-}\textbf{b})}_\text{o}$, the spectrum weight for \textbf{b}-\textbf{g} reaches its maximum values at $\Gamma^z=\pm3$, showcasing two peaks when $J_z=0.6$ {[Fig.~\ref{fig04}(e)]}.
This indicates that CS exhibits larger quantum fluctuations and is more ``soft" at small $J_z$.
However, these peaks vanish due to the dominant $H_\text{d}$ at large $J_z$ {[Figs.~\ref{fig04}(a,c)]}.

\subsection{Wave packs}
%one-dimensional picture
We have demonstrated the existence of CS and elucidated its movement, akin to the movement of strings in frustrated systems~\cite{dqcp01, zhou01, string_tri}.
It is important to note that such dimensionality reduction is not a result of the aspect ratio chosen in $2$D DMRG.
Even with larger $L_y$ {[e.g., $L_y=8$ in Appendix~\ref{AppA}]}, the result is in good agreement with the effective model.
As the string approaches the edge columns, it can ``feel" the repulsion interactions, particularly when the distance is shorter than the characteristic distance $d_\text{ec}$. This phenomenon is analog to the entropy-reduced force observed at finite temperatures~\cite{Baxter}.

\begin{figure}[t]
\centering
\includegraphics[width=0.99\linewidth]{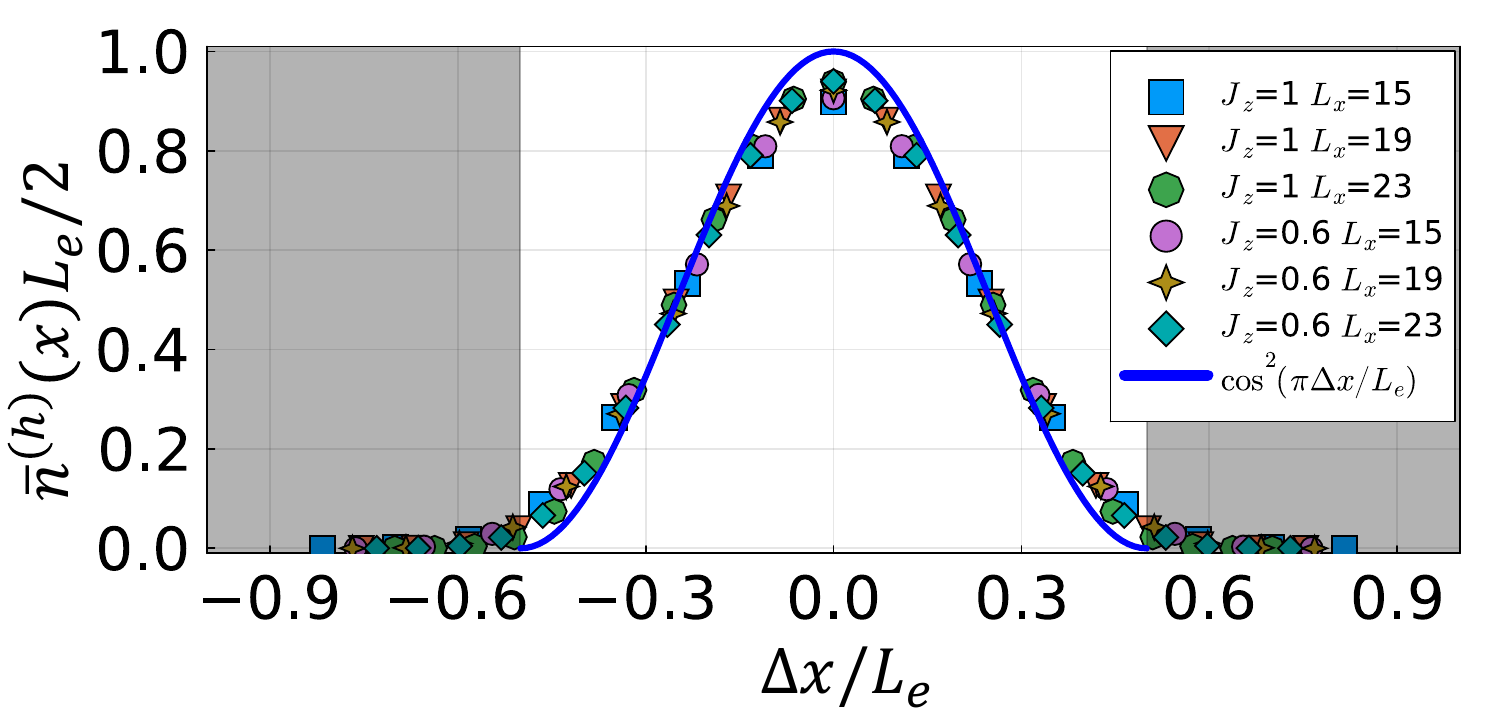}
\caption{The rescaled average hole density in a $L_x \times 6$ cylinder with the length $L_x = 15$, $19$, $23$, as calculated by $2$D DMRG, is presented as a function of the rescaled displacement in the $t$-$J_z$ model at $J_z=1$ and $0.6$.
The gray area indicates boundary-interactive regions. The blue solid line gives a function of $\cos (\pi \Delta x / L_\text{e})$ in the white area.
%The mess grey dashed lines demonstrate the schematical picture of the movement (green arrows) of the string due to vibrations. 
 }\label{fig07}
\end{figure}

To simplify, we treat CS as a ``hardcore" particle with a finite core radius of $L_\text{s} / 2$, which could be smaller than $d_\text{ec}$.
For the lowest-energy translation mode, the distribution of the average hole density is readily derived as
\begin{equation}
\bar{n}^{(\text{h})} (x) = \frac{2}{L_\text{e}} \cos^2 \left(\frac{\pi \Delta x}{L_e}\right)\, ,\nonumber
\end{equation}
where $\Delta x = x - x_\text{c}$ represents the displacement from the center $x_\text{c} = (L_x + 1) / 2$, and $L_\text{e} = L_x - L_\text{s}$ is the effective length for the CS motion.
After rescaling, as shown in Fig.~\ref{fig07}, the curves $\bar{n}^{(\text{h})} (x)$ at different values of $L_x$ and $J_z$ can be collapsed onto a single curve with a tunable parameter $L_\text{e}$.

%Energy
From the CS scenario, we can immediately propose a scaling ansatz for the groundstate energy of a CS in the AF background, i.e.,
\begin{equation}
\nonumber
    E=-NJ_z - 2 t_\text{s} \cos \left( \frac{\pi}{L_\text{e}} \right) + \beta\left(L_y\right)\, ,
    \label{eq3}
\end{equation}
where the first leading term signifies the energy contribution from the AF background, the second term represents the kinetic energy arising from the translation of CS characterized by an effective hopping amplitude $t_\text{s}$, and the last term accounts for the internal energy associated with the vibration of CS dependent of $L_y$.

\begin{figure}[h]
\centering
\includegraphics[width=0.99\linewidth]{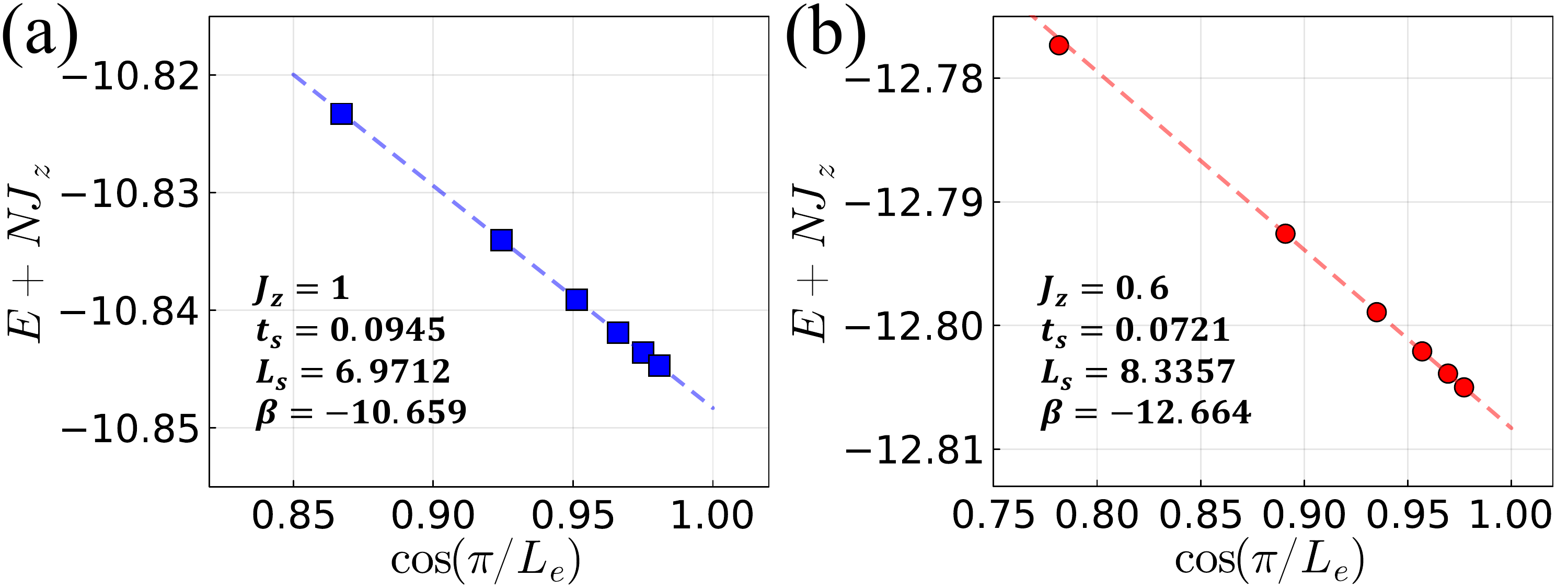}
\caption{The residual energy of the $t$-$J_z$ model calculated by $2$D DMRG at (a) $J_z=1$ and (b) $0.6$ in a $L_x \times 6$ cylinder with the length $L_x = 13$, $\cdots$, $23$. Two dashed lines give the best linear fitting.  
 }\label{fig08}
\end{figure}

To validate the ansatz, we use $2$D DMRG to compute the groundstate energy at different values of $L_x$ when $L_y=6$.
In Fig.~\ref{fig08}, after subtracting the leading term, the residual energy manifests an excellent function of $\cos(\pi/L_e)$.
Correspondingly, the core radii are around $3.5$ at $J_z=1$ with slight discrepancies, which is close to the predicted $L_y/2=3$ by Zaanen.
At lower $J_z$, the increased core radius reflects the softening of the CS.
Most importantly, the effective hopping amplitude $t_s$ can be extracted from the energy scaling, and its magnitude is $\sim 0.1$ which is strikingly close to the energy scale of the room temperature when considering the recognized model parameters for cuprates~\cite{RevModPhys.66.763}. 

\subsection{Pinning field}
\begin{figure}[t]
\centering
\includegraphics[width=0.99\linewidth]{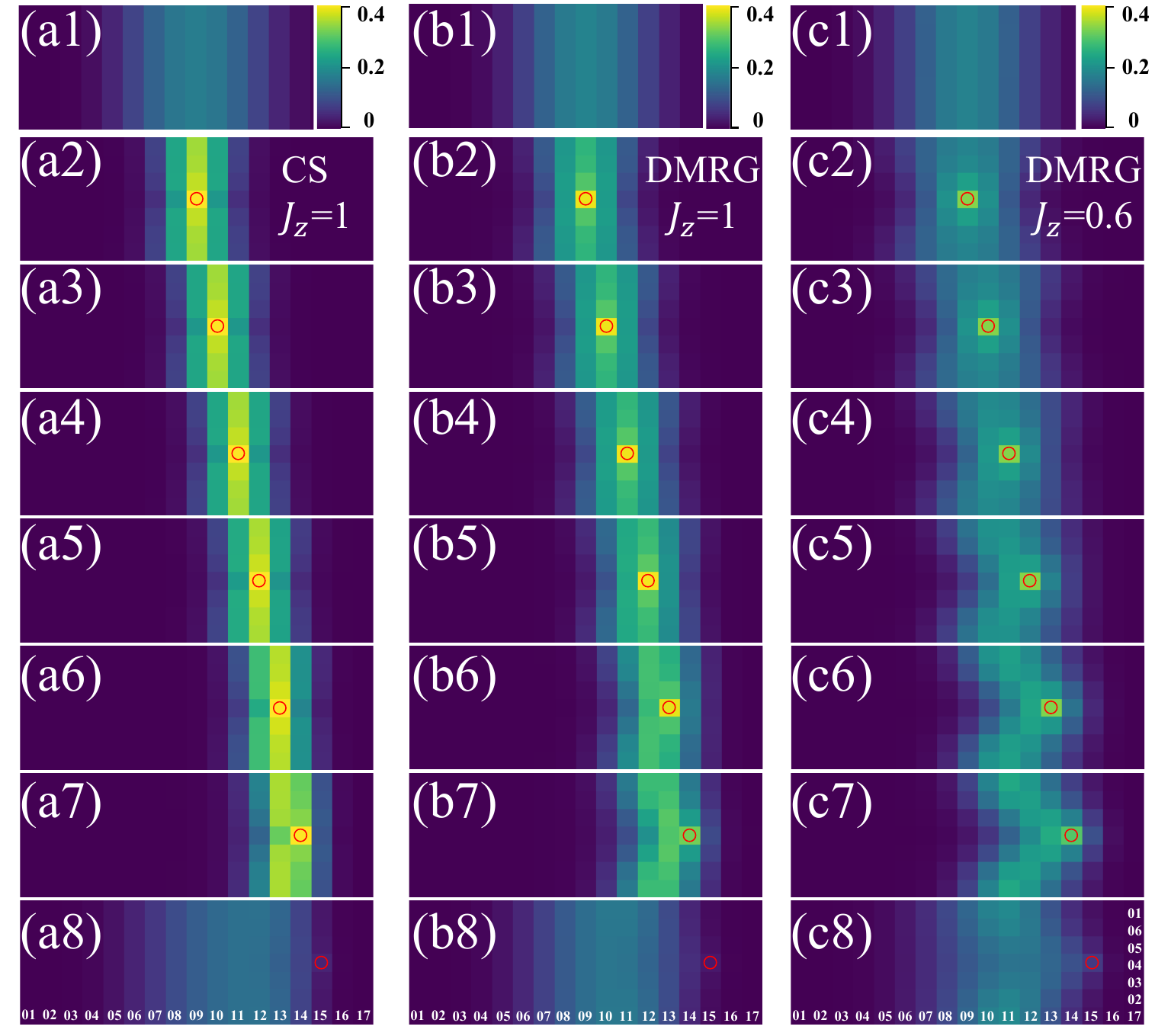}
\caption{The distribution of the local hole density $n^{(\text{h})} (x,\ y)$ in a $17\times6$ cylinder, before (a1,b1,c1) and after (a2-a8,b2-b8,c2-c8) applying a pinning field with the strength $V_\text{p}=0.5$ at different sites (red circles).
A comparison is made between (a) the ED results of the effective model~\eqref{effHam} at $J_z=1$, and (b,c) the DMRG results of the $t$-$J_z$ model~\eqref{eq1} at (b) $J_z=1$ and (c) $0.6$.
 }\label{fig09}
\end{figure}

Following the Zaanen's description, the quantum strings in cuprates are ``rivers of charge"~\cite{Zaanen01}, or maybe alternatively rephrased as ``holes moving in the river."
In our new scenario, three distinct CPs can move in the CS.
Next, we can introduce a small pinning field $-V_\text{p} (1 - n_{i,\uparrow} - n_{i,\downarrow})$ at site-$i$ with the field strength $V_\text{p}=0.5$ to tweezer the CS {[Figs.~\ref{fig09}(a2,b2,c2)]}.

As illustrated in Fig.~\ref{fig09}, both the effective model and the $t$-$J_z$ model demonstrate that the pinning field is capable of dragging CS to the vicinity of the right edge column.
In these cases, the field strength is much larger than the kinetic energy of the CS, which is approximately $2 t_\text{s}$.
At large $J_z$ {[Figs.~\ref{fig09}(a,b)]}, the CS bends when its distance from the edge column falls within the range of the characteristic distance $d_\text{ec} > L_{s} / 2$, indicating a deviation from a rigid ``hardcore" behavior.
When the pinning field shifts to the last third $x$-position {[Figs.~\ref{fig09}(a8,b8,c8)]}, it is unable to pin the CS because the deformation energy of the CS becomes larger than the pinning energy.
At smaller $J_z=0.6$, the string can feel the edge column earlier {[Fig.~\ref{fig09}(c4)]}, which reflects a larger core radius, thereby supporting the notion of the CS softening in Fig.~\ref{fig08}.

\subsection{Partially-filled CS}
\begin{figure}[t]
	\centering
	\includegraphics[width=0.99\linewidth]{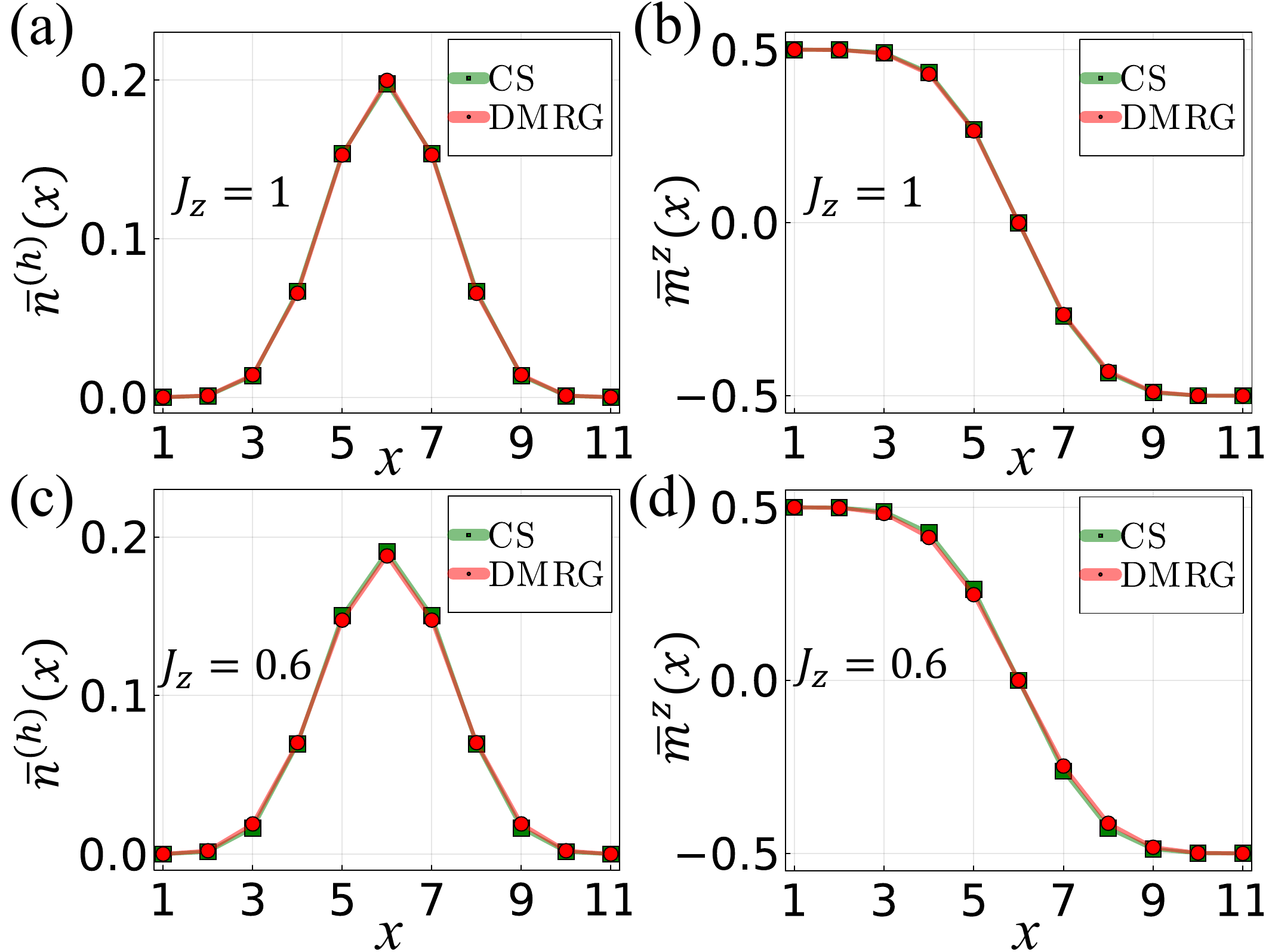}
	\caption{{The distribution of the average hole density $\bar{n}^{(h)}(x)$(left panel) and average magnetic moment $\bar{m}^z(x)$(right panel) alone x-direction in a $11 \times 6$ cylinder with four holes. The notation is the same as Fig.\ref{fig05}}}\label{fig11}
\end{figure}

In %the fully-filled CS
{the filled CS}, the \textbf{r} and \textbf{b}-particles are excited from the \textbf{g}-\textbf{g} pair so that we can immediately obtain the conservation relation $n^{\textbf{(r)}}=n^{\textbf{(b)}}$. However, if the number of holes is less than $L_y$, the so-called partially-filled CS may have more \textbf{r}-particles than \textbf{b}-particles, i.e. violation of conservation.
{For example, $\nu =2/3$ for $L_y=6$ in Fig.~\ref{fig11}.}

As analysis of %the fully-filled CS
{the filled CS}, we first study the distribution of the hole density and magnetic moment.
In comparison with the DMRG method {[Fig.~\ref{fig11}]}, both observables demonstrate that the {colored string} still quantitatively works well for partially-filled CS at large $J_z$. When $J_z$ becomes smaller, the wave pack becomes broad, which indicates the softening of the quantum string is mainly related to the strength of local antiferromagnetic order but less relevant to the hole-doping density. 

\begin{figure}[t]
	\centering
	\includegraphics[width=0.99\linewidth]{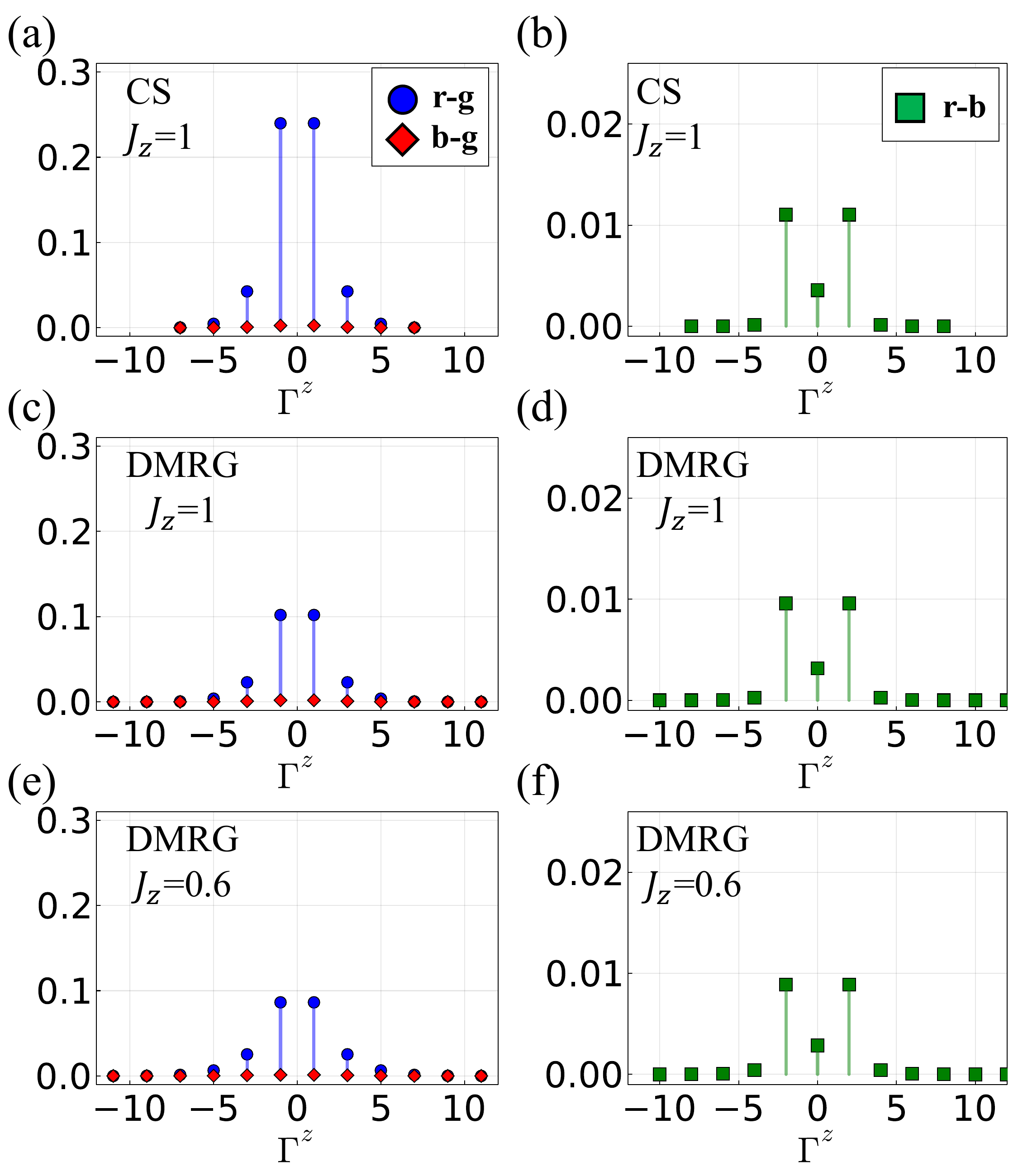}
	\caption{The spectrum weight $\Lambda \left(c, c', \Gamma^z \right)$ as a function of the spin field $\Gamma^z$ in a $11 \times 6$ cylinder with four holes. The notation is the same as Fig.\ref{fig06}
	}\label{fig12}
\end{figure}

Although the profile of the distribution does not present much difference, as the fingerprint, the spectrum weight of the effective spin field provides more insight into the interplay between CPs and effective spin fields. As shown in Fig.~\ref{fig12}, a small $|\Gamma_{z}|$ also exhibits a higher distribution. However, compared to %the fully-filled CS
{the filled CS}, the {violation} of conservation results in the \textbf{r}-particle to take the main effect. Then, the spectrum weight for \textbf{b}-\textbf{g} is totally suppressed. However, the decrease in the strength of the \textbf{r}-\textbf{b} pair is not severe, so that the possible superconductivity relation can be built through the following process:${\uparrow\downarrow\downarrow\uparrow}/{\uparrow\circ\circ\uparrow}\rightarrow{\uparrow\downarrow\circ\circ}/{\uparrow\downarrow\uparrow\uparrow}$.

\section{conclusion and discussion}\label{sec5}
A bottom-up effective theory has been developed to explain the striped phase of the hole-doped $t$-$J_z$ model, where the quantum string in the antiferromagnetic background is colored with three distinct particles, beyond the existing description of sublattice parity order~\cite{Zaanen01, Fabian}. Meanwhile, the colored string is also different from the {``phase string"~\cite{RevModPhys.66.763, wenzhengyu}, which is recently found to play a critical role in forming $d$-wave superconductivity~\cite{phase_string1,phase_string2}}.
Furthermore, the application of a pinning field, such as the tweezer light in an optical lattice~\cite{Young_2022}, enables visualization and manipulation of the quantum string.
These findings are semi-quantitatively validated through large-scale $2$D-DMRG calculations.

%In the quantum material, the pinning field may be added \textit{via} the insertion of impurity on the substrate. In the optical lattice, the local field can be implemented with optical tweezer light.

Our effective theory may be viewed as a $1$+$1$D gauge theory, in which color particles can interact with a $U(1)$ gauge field. Off-diagonal interactions induce vibrations in the string.
The terms along the string facilitate the creation and movement of spinons and dual-holes, which is crucial for electron pairing.
In this work, we only want to show that the quantum string is the intrinsic physics picture of the stripe phase, otherwise the basis of {colored string} will be invalid.
{The discussion at the end of partially-filled CS section has already shown some hints about the superconducting correlation.}
Therefore, our effective theory not only provides a more quantitative description of the quantum string, but also may establish a direct connection to superconductivity \cite{yayu_2023}.

Indeed, this effective theory can be readily extended to many directions, e.g., $t$-$t'$-$J_z$ model and $t$-$J$ model.
{Regarding the $t$-$J$ model, adding spin-flipping terms into the $t$-$J_z$ model, enhances the flexibility and quantum nature of the colored string from a microscopic perspective.
This enhancement may increase the repulsion energy of two adjacent strings and strengthen the stripe orders.
Our recent work~\cite{pm} further indicates that this enhancement can boost spinon-pair exchange, i.e., \textbf{r} particle, which further promotes $d$-wave pairing.}

%Additionally, it can be used to investigate the electron-doped case where the hole is replaced by the doublon.
%Thus, we believe that our work may lead to a significant breakthrough in the comprehension of high-$T_c$ superconductivity mechanisms.

%From the viewpoint of the field theory, three types of color particles can be taken as different matter fields, and the effective spin field behaves like a $U(1)$ gauge field.
%Therefore, our effective theory may be taken as a strongly interactive gauge theory with matter fields.
%From the effective theory, we can find all the off-diagonal interactions join into shaking the string, but only the interactions along the string produce and move the spinon and dual-hole which may play a critical role in the superconductivity.
%Definitely, such discussion is not strict but may intrigue the interest of theoretical physicists.

\section{Acknowledgement}
We would like to thank Hao~Ding, Shi-Ping~Feng, Jin~Zhang, Shi-Wei~Zhang, and Yuan-Yao~He for many helpful discussions. This paper is also in memory of Prof.~Peter Fulde and Prof.~Jan Zaanen \cite{phillips2024smoke}, who contributed a lot to the quantum strings. 
S. J.~H. acknowledges funding from MOST Grant No.~2022YFA1402700, NSFC No.~U2230402, NSFC Grant No.~12174020.
X.-F.~Z. acknowledges funding from the National Science Foundation of China under Grants No.~12274046, No.~11874094, No.~12147102, and No.~12347101, the Chongqing Natural Science Foundation under Grant No.~CSTB2022NSCQ-JQX0018, the Fundamental Research Funds for the Central Universities Grant No.~2021CDJZYJH-003, and the Xiaomi Foundation/Xiaomi Young Talents Program.

\appendix\label{APPENDIX}

\renewcommand\thefigure{S\arabic{figure}}
\setcounter{figure}{0}

\renewcommand{\thetable}{T\arabic{table}}
\setcounter{table}{0}
%\section*{Appendix}

\section{Phase transition}\label{AppA1}
To show the transition from the hole cluster state to the string state with a $\pi$-phase shift, we use $2$D DMRG to calculate the energy of two low-energy states as a function of the parameter $J_z$.
The energy values are presented in the Table~\ref{energy0pi}.
Then, we can find that the ground state exhibits a preference for the $\pi$-phase string state when $J_z \le 3.8$, while the hole cluster state has a lower energy when $J_z \ge 3.9$. Therefore, we estimate that the transition point occurs at $J_{z,c} \simeq 3.85$.

\begin{table}[h]
    \centering
    \begin{tabular}{|c|c|c|}
	\hline
	$J_z$ & Hole cluster & $\pi$-phase string \\
	\hline
    	$3.5$ & $-235.3558981$ & $-235.6769177$ \\
	\hline
	$3.6$ & $-241.8920491$ & $-242.1266182$ \\
	\hline
	$3.7$ & $-248.4342094$ & $-248.5815498$ \\
	\hline
	$3.8$ & $-254.9844699$ & $-255.0416582$ \\
	\hline
	$3.9$ & $-261.5423043$ & $-261.5068928$ \\
	\hline
	$4$    & $-268.1071546$ & $-267.9772034$ \\
	\hline
    \end{tabular}
    \caption{The energy values for two competing states calculated by $2$D DMRG with the bond dimension $\chi = 2,048$.}
    \label{energy0pi}
\end{table}

\section{Colored string}\label{AppCS}

\subsection{Onsite energy of color particles}\label{App:A1}

The \textit{onsite} energy of CPs is only related to the violation of AF bonds in the $x$-axis, as outlined in Table~\ref{tabel1}.
%It can be observed that these operators are always bosonic, given that an even number of fermionic operators are involved in their definitions.
%It is noteworthy that the braiding of two spinons are prohibited in the effective model, despite that fact that spinons are typically regarded as anyons with fractional statistic~\cite{}. Consequently, they can be effectively seen as bosons.
\begin{table}[h!]
    \centering
    \begin{tabular}{|c|c|c|c|}
    \hline
     Particle &  Color  & Configuration & Onsite energy \\
         \hline
      Spinon   & \textbf{r} & $\uparrow\downarrow\downarrow\uparrow$ and $\downarrow\uparrow\uparrow\downarrow$ & $J_z/2$\\
      \hline
       Holon  & \textbf{g} & $\uparrow\!\!\circ\!\!\downarrow$ and $\downarrow\!\!\circ\!\!\uparrow$ & $J_z$\\
      \hline
       Dual-hole  & \textbf{b} & $\uparrow\!\!\circ\circ\!\!\uparrow$ and $\downarrow\!\!\circ\circ\!\!\downarrow$ & $3J_z/2$\\
       \hline
    \end{tabular}
    \caption{The onsite energy of CPs under consideration.}
    \label{tabel1}
\end{table}

\subsection{String vibration}\label{App:A2}

\begin{figure}[b]
\centering
\includegraphics[width=0.99\linewidth]{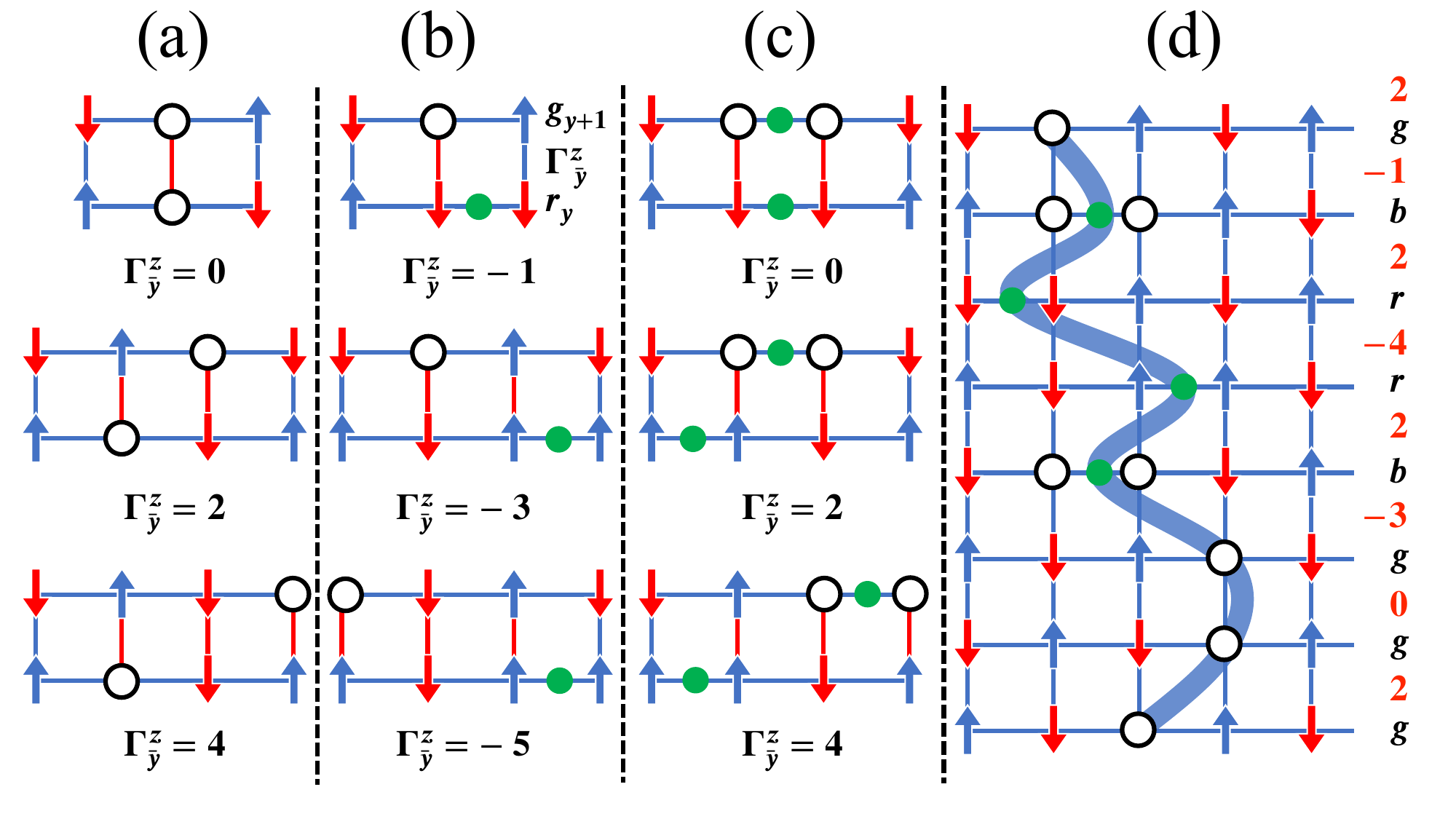}
\caption{Effective spin fields for showing the displacements between CPs for (a) $c_y$-$c_{y+1}=\textbf{g}$-$\textbf{g}$, (b) \textbf{g}-\textbf{b}, and (c) \textbf{r}-\textbf{b}.
A red link highlights a violated AF bond with an energy cost of $J_z/2$, and a green dot indicates the position of \textbf{r} or \textbf{b}.
(d) A mapping of a electron configuration to a CS (blue line).
 }\label{fig02}
\end{figure}

The hole hopping along either the $x$-axis or $y$-axis changes the \textit{filling status} of CPs in two adjacent rows.
We begin by assuming that the location of the CP in row-$1$ is fixed, and then observe the vibration of the CS.
For example, in the process
\begin{equation}
\textbf{g}\text{-}\textbf{g}: \begin{array}{cccc}
\uparrow & \circ & \downarrow & \uparrow\\
\hline
\uparrow & \downarrow & \circ & \uparrow\\
\end{array}
\Longrightarrow
\textbf{r}\text{-}\textbf{b}: \begin{array}{cccc}
\uparrow & \circ & \circ & \uparrow\\
\hline
\uparrow & \downarrow & \downarrow & \uparrow\\
\end{array}\ ,\nonumber
\label{picture-gg-rb}
\end{equation}
adjacent rows in the initial configuration on the left side are filled with two holons, forming a \textbf{g}-\textbf{g} pair, while in the final configuration on the right side, they contain a dual-hole and a spinon, forming an \textbf{r}-\textbf{b} pair.
According to Table~\ref{tabel1}, we find that the transition from the \textbf{g}-\textbf{g} pair to the \textbf{r}-\textbf{b} pair does not entail any onsite energy cost.
%\change{Later on, we will find these color particles can make colored string become `colorful'.}{Therefore, it is not necessary to consider onsite energy in the effective theory here. Then, the effective theory only needs to consider two parts: (1) the \textit{diagonal interactions} related to the distance between color particles; (2) the \textit{off-diagonal interactions} which deform the string, produce the \textbf{r-g} pair and move the color particles along the strings. Furthermore, the two degeneracy states of the antiferromagnetic domain can be labeled as AM-$A$ and AM-$B$ depending on the spin-up locating at which sublattice. Then, it is obvious that the $\pi$-phase shift just happens at the position of the color particle.}
%Therefore, it is not necessary to consider onsite energy in the effective theory here.

The effective theory for hole hopping processes can be classified into two categories: (1) \textit{diagonal interactions}, which are influenced by the distance between CPs, and (2) \textit{off-diagonal interactions}, which deform the CS, produce the \textbf{r}-\textbf{b} pair, and move CPs along the CS.
Then, it is obvious that the $\pi$-phase shift happens at the CP location in each row.
%Furthermore, the two degeneracy states of the antiferromagnetic domain can be labeled as AM-$A$ and AM-$B$ depending on the spin-up locating at which sublattice.

%Gauge Field and Diagonal interaction
\textit{Diagonal interactions} - The violation of the AF bonds along the $y$-axis results in the diagonal interaction among CPs, indicated by the red links in Fig.~\ref{fig02}(a-c).
%This means that configurations with larger distances have higher energy values and are thus significantly discouraged in the ground state, reminiscent of the Nagaoka mechanism.
%Then, the sign of $\Gamma^z_{\bar{y}}$ depends on the relative direction between CPs, and it does not affect the interaction.
%The same features can also be found in the interactions between different species of CPs, as demonstrated in Fig.~\ref{fig02}(b,c).
In Fig.~\ref{fig02}(d), the configuration of electrons in a cylinder can be mapped to a CS, which consists of three distinct species of CPs and is coupled to the effective spin field located at dual sites of the CS.
All diagonal interactions $V_{c,c'}$ are listed in Table~\ref{tabel2}.
%After calculating all interactions listed in Table~\ref{tabel2}, we can derive the diagonal energy of a CS, i.e.,
%\begin{eqnarray}
%\nonumber
%H_\text{d} = \sum_y \sum_{c, c'} V_{c,c'} \left( \lvert \Gamma^z_{\bar{y}} \rvert \right) n^{(c)}_y n^{(c')}_{y+1}\ .
%\label{}
%\end{eqnarray}
%\begin{table}[h]
%    \centering
%    \begin{tabular}{|c|c|c|c|c|c|c|}
%    \hline
%     Type &  $\Gamma^z$  & Energy &  $\Gamma^z$  & Energy &  $\Gamma^z$  & Energy \\
%         \hline
%      \textbf{g}-\textbf{g}   & $0$ & $J_z/2$ & $\pm 2$ & $J_z$ & $\pm 2Z$ & $(Z+1)J_z/2$\\
%         \hline
%      \textbf{r}-\textbf{g}   & $\pm 1$ & $J_z/2$ & $\pm 3$ & $J_z$ & $\pm (2Z+1)$ & $(Z+1)J_z/2$\\
%         \hline
%      \textbf{g}-\textbf{b}   & $\pm 1$ & $J_z$ & $\pm 3$ & $3J_z/2$ & $\pm (2Z+1)$ & $(Z+2)J_z/2$\\
%         \hline
%      \textbf{r}-\textbf{b}   & $0$ & $J$ & $\pm 2$ & $J_z$ & $\pm 2Z$ & $(Z+1)J_z/2$\\
%         \hline
%      \textbf{r}-\textbf{r}   & $0$ & $0$ & $\pm 2$ & $J_z/2$ & $\pm 2Z$ & $ZJ_z/2$\\
%         \hline
%      \textbf{b}-\textbf{b}  & $0$ & $J_z$ & $\pm 2$ & $3J_z/2$ & $\pm 2Z$ & $(Z+2)J_z/2$\\
%       \hline
%    \end{tabular}
%    \caption{The strength of the diagonal interaction $V_{\textbf{c}, \textbf{c}'} \left(\Gamma^z_l\right)$.}
%    \label{tabel2}
%\end{table}
\begin{table}[h]
    \centering
    \begin{tabular}{|c|c|c||c|c|c|}
    \hline
     Conf.: $c$-$c'$ &  $\lvert \Gamma^z_{\bar{y}} \rvert$  & $V_{c, c'} / J_z$ & Conf.: $c$-$c'$ &  $\lvert \Gamma^z_{\bar{y}} \rvert$  & $V_{c, c'} / J_z$\\
         \hline
      \textbf{g}-\textbf{g} & $2 \mathbb{N}$ & $(\mathbb{N}+1)/2$ & \textbf{r}-\textbf{r} & $2\mathbb{N}$ & $\mathbb{N}/2$\\
         \hline
      \textbf{r}-\textbf{g} & $2\mathbb{N}+1$ & $(\mathbb{N}+1)/2$ & \textbf{r}-\textbf{b} & $2\mathbb{N}$ & $(\mathbb{N}+1)/2$\footnote{When the natural number $\mathbb{N}=0$, $V_{c, c'}=J_z$ does not follow the formula.}\\
         \hline
      \textbf{g}-\textbf{b} & $2\mathbb{N}+1$ & $(\mathbb{N}+2)/2$ & \textbf{b}-\textbf{b} & $2\mathbb{N}$ & $(\mathbb{N}+2)/2\ \,$\\
         \hline
    \end{tabular}
    \caption{Strengths of non-zero diagonal interactions $V_{c, c'} \left( \lvert \Gamma^z_{\bar{y}} \rvert \right)$.}
    \label{tabel2}
\end{table}

\begin{figure}[b]
\centering
\includegraphics[width=0.99\linewidth]{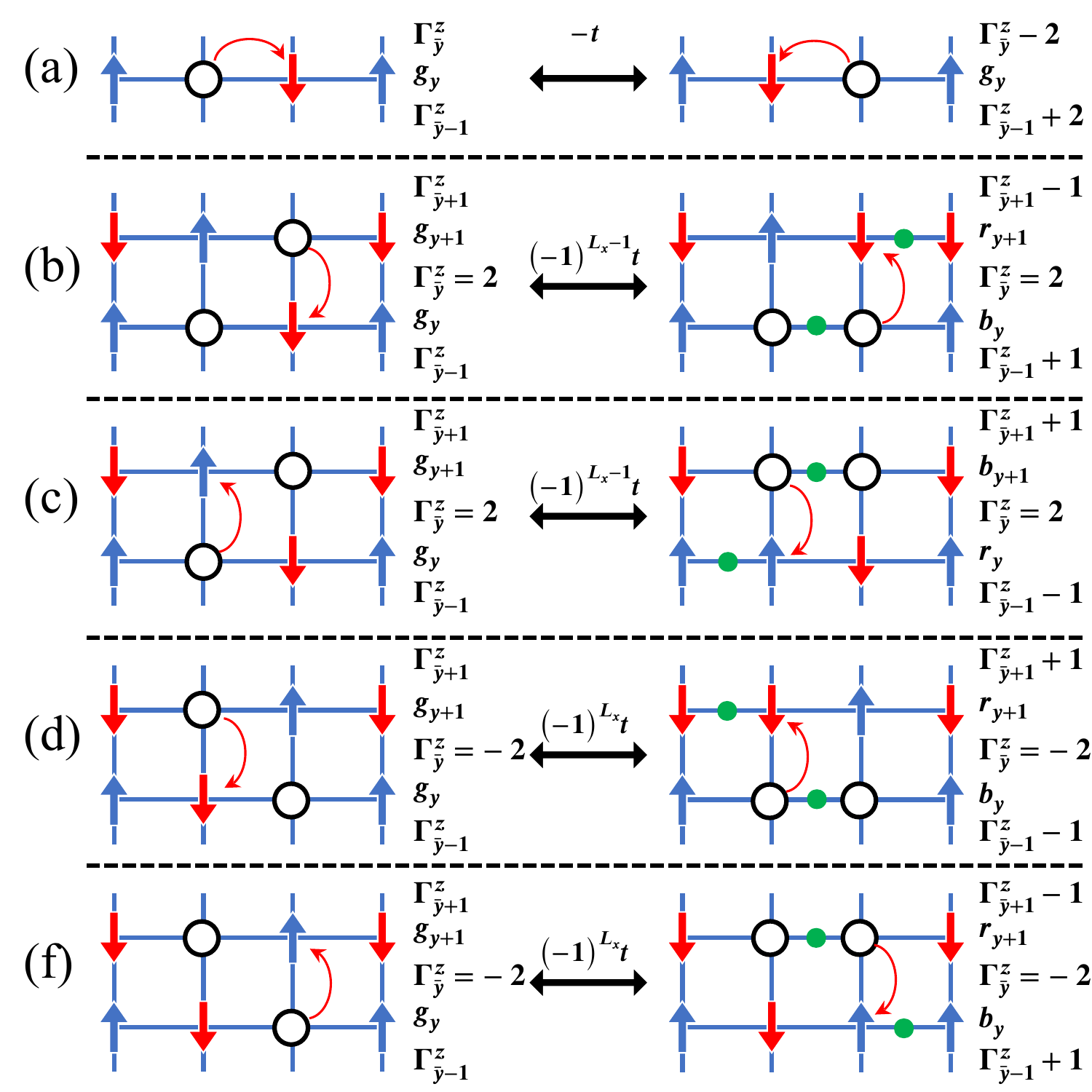}
\caption{Off-diagonal interactions update the configurations of CS \textit{via} hole hoppings (red arrow) (a) along the $x$-axis, and (b-d) along the $y$-axis for the process \textbf{g}-\textbf{g} $\Longrightarrow$ \textbf{r}-\textbf{b} and reversion.
%Red arrows indicate the hopping processes of holes.
 }\label{fig03}
\end{figure}

\begin{figure}[t]
\centering
\includegraphics[width=0.99\linewidth]{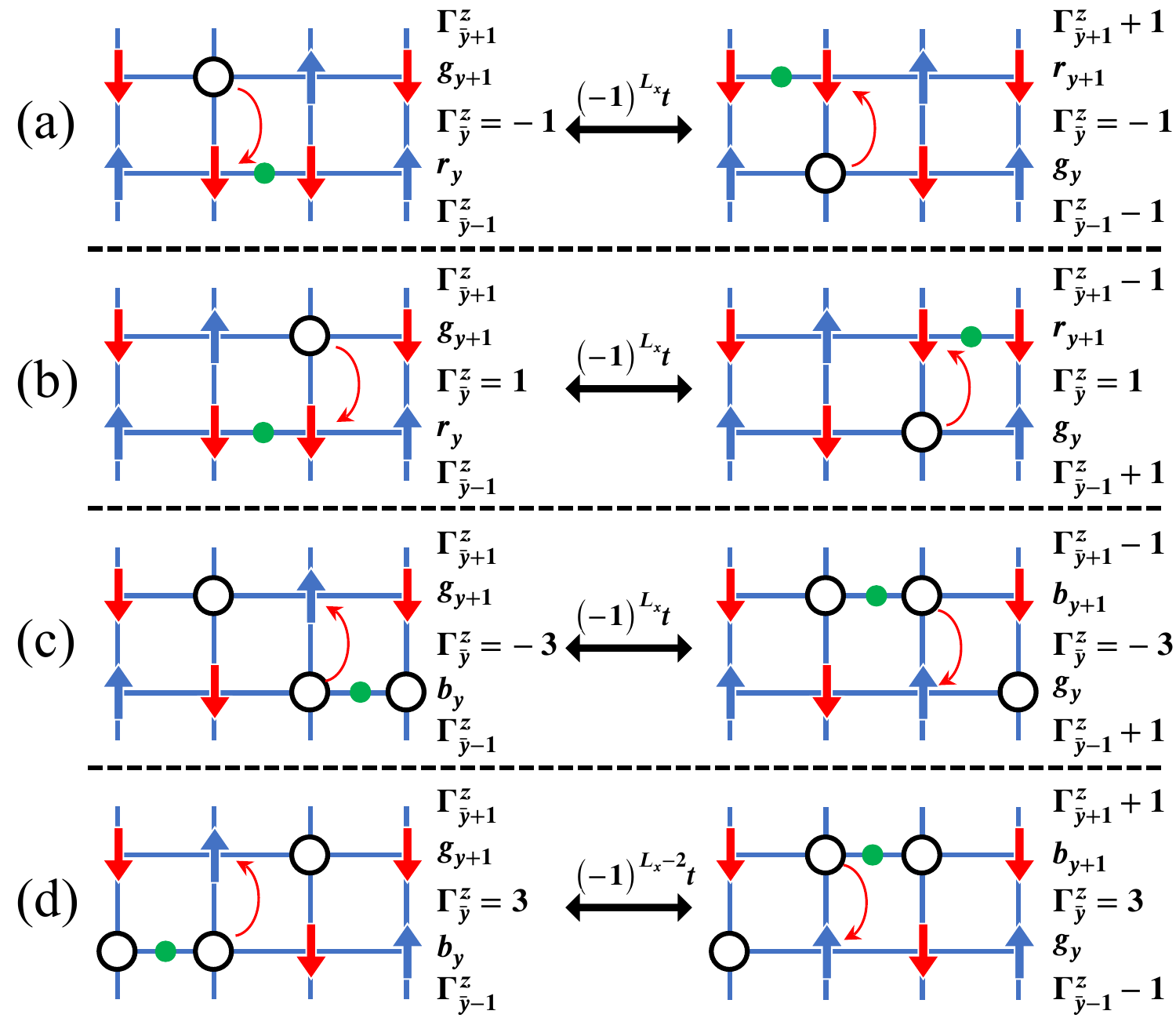}
\caption{
Other off-diagonal interactions update the configurations of CS \textit{via} hole hoppings (red arrow) for the processes (a,b) \textbf{r}-\textbf{g} $\Longrightarrow$ \textbf{g}-\textbf{r}, and (c,d) \textbf{b}-\textbf{g} $\Longrightarrow$ \textbf{g}-\textbf{b}.
 }\label{fig04}
\end{figure}

%Dynamical Process and Off-diagonal interaction
%\textit{Off-diagonal interactions}---The shape of the CS changes through the hopping processes of holes, which contributes to the vibrational energy.
%Same as Zaanen's string, when a hole in row-$y$ hops to the left or the right, the effective spin fields in neighboring dual rows $(\bar{y} - 1)$ and $\bar{y}$ change as well, as shown in Fig.~\ref{fig03}(a).
%Defining the ``shift operator" $\Gamma^+_{\bar{y}}$ ($\Gamma^-_{\bar{y}}$) which raises (lowers) spin field as $\Gamma^\pm_{\bar{y}} \ket{\Gamma^z_{\bar{y}}} = \ket{\Gamma^z_{\bar{y}} \pm1}$, we can derive the operator representation of the interaction, given by
%\begin{equation}
%H_\text{o}^{(\textbf{g})} = -t \sum_y \left[ \left(\Gamma^+_{\bar{y}}\right)^2 n^{(\textbf{g})}_y \left(\Gamma^-_{\bar{y}-1} \right)^2 + \textrm{h.c.}\right]\, .
%\label{eqn04}
%\end{equation}
%Then, we can establish a connection with Zaanen's string by noting that the spin field reduces to the one for the spin-$1$ through the relations $\left( \Gamma^\pm_{\bar{y}} \right)^2 \leftrightarrow S^\pm_{\bar{y}}$ and $\Gamma^z_{\bar{y}} \leftrightarrow 2 S^z_{\bar{y}}$.

\textit{Off-diagonal interactions}---Same as Zaanen's string, when a hole in row-$y$ hops to the left or the right, the effective spin fields in neighboring dual rows $(\bar{y} - 1)$ and $\bar{y}$ change as well [Fig.~\ref{fig03}(a)].
The hopping processes of holes along the $y$-axis bring about more complex changes to the CS.
For instance, referring to Fig.~\ref{fig03}(b-f), when the spin field $\Gamma^z_{\bar{y}}$ between \textbf{g}-\textbf{g} equal to $\pm2$, they can transform to the \textbf{r}-\textbf{b} pair.
Interestingly, the spin field in the middle remains unaffected, whereas the upper and lower ones alter through a form of spin exchange interaction.
Then, the sign of the transition amplitude depends on the sign of the spin field $\Gamma_{\bar{y}}^z$, behaving akin to a gauge field.
The corresponding Hamiltonian for $\Gamma^z_{\bar{y}} = \pm2$ can be explicitly written as
\begin{equation}
\begin{split}
H^{(\textbf{r}\text{-}\textbf{b} / \textbf{g}\text{-}\textbf{g})}_\text{o} =\pm t \sum_y \left(\Gamma^\mp_{\bar{y}+1} \text{r}_{y+1}^\dag \text{b}_y^\dag \Gamma^\pm_{\bar{y}-1} + \Gamma^\pm_{\bar{y}+1} \text{b}_{y+1}^\dag \text{r}_y^\dag \Gamma^\mp_{\bar{y}-1}\right) \text{g}^{\phantom{\dag}}_{y+1} \text{g}^{\phantom{\dag}}_y\, .\nonumber
\end{split}
\end{equation}

Meanwhile, we find that the spinon and dual-hole can switch places with the holon when the corresponding middle spin field equal to $\Gamma^z_{\bar{y}} = \pm1$ and $\pm3$ [Fig.~\ref{fig04}], respectively.
The influence on the effective spin fields is the same type as the generation of the \textbf{r}-\textbf{b} pair.
However, their sign of hopping is independent of $\Gamma^z_{\bar{y}}$ and the corresponding Hamiltonians are
\begin{eqnarray}
\begin{split}
H^{(\textbf{r}\text{-}\textbf{g}/\textbf{g}\text{-}\textbf{r})}_\text{o} &= -t \sum_y \Gamma^\mp_{\bar{y}+1} \text{r}_{y+1}^\dag \text{r}^{\phantom{\dag}}_y \text{g}_y^\dag \text{g}^{\phantom{\dag}}_{y+1} \Gamma^\pm_{\bar{y}-1}\, ,
\hspace{10px} \Gamma_{\bar{y}}^z=\pm1\, .\\
\nonumber
H^{(\textbf{b}\text{-}\textbf{g}/\textbf{g}\text{-}\textbf{b})}_\text{o} &= -t \sum_y \Gamma^\pm_{\bar{y}+1} \text{b}_{y+1}^\dag \text{b}^{\phantom{\dag}}_y \text{g}_y^\dag \text{g}^{\phantom{\dag}}_{y+1} \Gamma^\mp_{\bar{y}-1}\, , \hspace{7px} \Gamma_{\bar{y}}^z=\pm3\, .
\end{split}
\end{eqnarray}

%Short Summary
Overall, the diagonal interactions limit the displacement between CPs, while the off-diagonal interactions both deform the CS and enable the motion of the spinons and dual-holes along the CS.
%all of which are summarized as the full Hamiltonian of the effective model of CS:
%Furthermore, without loss of generality, we take the $x$-coordinate of the first CP as the position of CS (e.g., $\mathcal{X}^{(\text{CP})}_1=3$ in Fig.~\ref{fig02}(d)), allowing for a one-to-one mapping of the real electron configurations during the exact diagonalization of the entire effective Hamiltonian of CS:
%\begin{equation}
%\begin{split}
%H^\text{CS}_\text{e} &= H_\text{d} + \left(H_\text{o}^{(\textbf{g})} + H^{(\textbf{r}\text{-}\textbf{b} / \textbf{g}\text{-}\textbf{g})}_\text{o} + H^{(\textbf{r}\text{-}\textbf{g}/\textbf{g}\text{-}\textbf{r})}_\text{o} + H^{(\textbf{b}\text{-}\textbf{g}/\textbf{g}\text{-}\textbf{b})}_\text{o} \right.\\
%&\left. + \textrm{h.c.}\right)
%\label{effHam}
%\end{split}\, .
%\end{equation}
%Drawing an analogy between the attachment of quasi-particles to a string and the stringing of pearls, we have chosen to refer to this innovative model as the 'Quantum Necklace Model'.

\begin{figure}[t]
\centering
\includegraphics[width=0.99\linewidth]{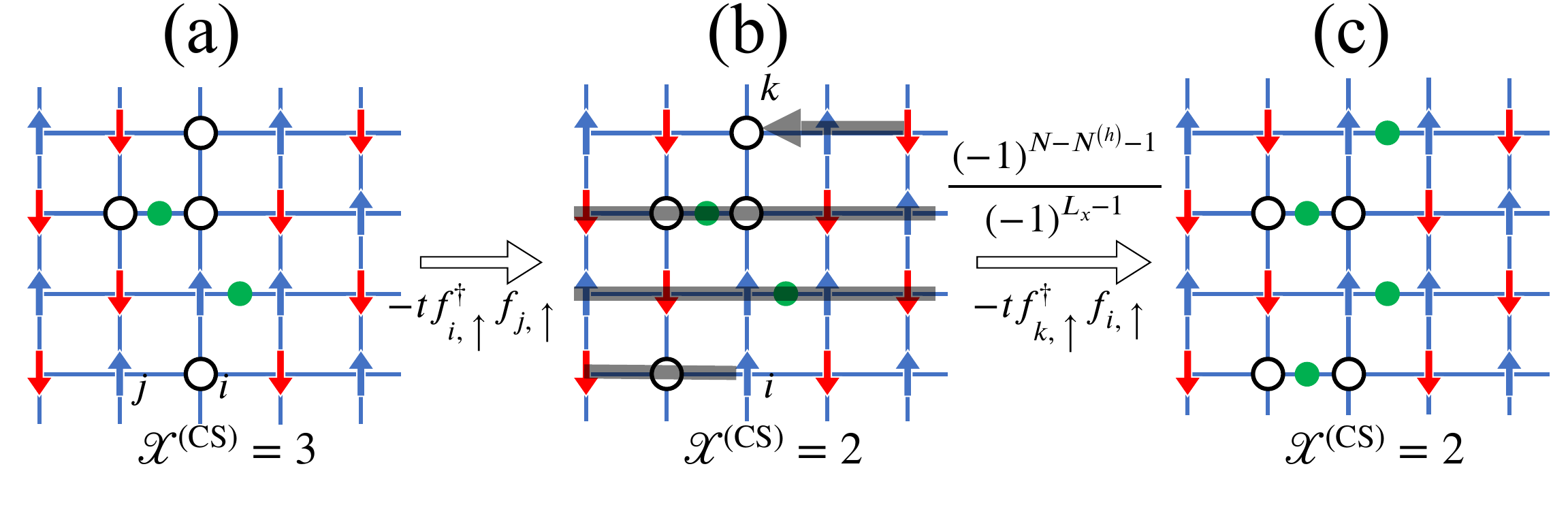}
\caption{Schematic images illustrate the configurations of CPs in the AF background.
(a)$\rightarrow$(b) shows a translation step of CS from the location (a) $\mathcal{X}^{(\mathrm{CS})}=3$ to (b) $\mathcal{X}^{(\mathrm{CS})}=2$ \textit{via} the hole hopping from site-$i$ to $j$ in row-$1$.
(b)$\rightarrow$(c) shows a change of the sign of the transition amplitude when the hole hops from site-$k$ to $i$.
The gray arrow in (b) indicates $1$D path for computing the sign.
%Here only off-diagonal processes are considered.
 }\label{fig05-06}
\end{figure}

\subsection{Translation}\label{App:A3}
After introducing a free CP in row-$1$, we can define the translation of the CS along the $x$-axis, as illustrated in Fig.~\ref{fig05-06}(a,b).
This allows CS to reach the leftmost or rightmost edges.
To simplify, we utilize $\mathcal{X}^{(\mathrm{CP})}_1$ to mark the position $\mathcal{X}^{(\text{CS})}$ of CS, that is, $\mathcal{X}^{(\text{CS})}_{\phantom{1}} \equiv \mathcal{X}^{(\mathrm{CP})}_1$.
The location of CP at any row can also be selected as $\mathcal{X}^{(\text{CS})}$, which does not affect the conclusion.
Definitely, the shift operator $\mathcal{X}^\pm$ describes the movement of CS in accordance with the relationship $\mathcal{X}^\pm \ket{\mathcal{X}^{(\mathrm{CS})}}=\ket{\mathcal{X}^{(\mathrm{CS})}\pm1}$.
Then, the corresponding operator representation of process Fig.~\ref{fig05-06}(a)$\rightarrow$(b) changes from the term for row-$1$ in Eq.~\eqref{eqn04} into
\begin{equation}
\nonumber
-t \left[ \left(\Gamma^+_{3/2} \right)^2 n^{(\textbf{g})}_1 \mathcal{X}^- \left(\Gamma^-_{1/2} \right)^2 + \textrm{h.c.}\right]\, ,
\end{equation}
along with the other off-diagonal interaction terms.
Meanwhile, we should be careful about the signs that emerge in the amplitudes resulting from the exchange of fermions along the $y$-axis.
In our work, given that both $N$ and $N^{(\text{h})}$ are even, like the case shown in Fig.~\ref{fig05-06}(b)$\rightarrow$(c), an additional prefactor $(-1)^{N-N^{(\text{h})}-1} = -1$ in the amplitude should been included.

\section{$2$D-DMRG benchmark for $L_y=6$}\label{AppB}
We show the truncation errors at different values of cylinder length $L_x$ and bond dimension $\chi$ when $L_y=6$ in Table~\ref{errors}, as the cylinder is partitioned into two equal-size segments in $2$D DMRG.
We find that $\chi=8,192$ can ensure the truncation errors below $10^{-5}$.
Moreover, the benchmark of the groundstate energy values is shown in Table~\ref{energy}, and the relative energy difference between $\chi=4,096$ and $8,192$ is $\sim 10^{-5}$.

\begin{table}[h]
    \centering
    \begin{tabular}{|c|c|c|c|}
	\hline
    	\multicolumn{4}{|c|}{$J_z=0.6$} \\
	\hline
        \multirow{2}*{$L_x$} & \multicolumn{3}{|c|}{$\chi$} \\
        \cline{2-4}
	& $2,048$ & $4,096$ & $8,192$ \\
	\hline
    	$11$ & $1.28\times10^{-4}$ & $4.28\times10^{-5}$ & $9.95\times10^{-6}$ \\
	\hline
    	$13$ & $1.13\times10^{-4}$ & $3.77\times10^{-5}$ & $8.91\times10^{-6}$ \\
	\hline
    	$15$ & $9.21\times10^{-5}$ & $3.11\times10^{-5}$ & $7.34\times10^{-6}$ \\
	\hline
    	$17$ & $7.77\times10^{-5}$ & $2.57\times10^{-5}$ & $5.98\times10^{-6}$ \\
	\hline
    	$19$ & $6.57\times10^{-5}$ & $2.13\times10^{-5}$ & $4.99\times10^{-6}$ \\
	\hline
    	$21$ & $5.60\times10^{-5}$ & $1.81\times10^{-5}$ & $4.26\times10^{-6}$ \\
	\hline
    	$23$ & $4.89\times10^{-5}$ & $1.61\times10^{-5}$ & $3.70\times10^{-6}$ \\
	\hline
    \end{tabular}\\
    \begin{tabular}{|c|c|c|c|}
    	\hline
    	\multicolumn{4}{|c|}{$J_z=1$} \\
	\hline
        \multirow{2}*{$L_x$} & \multicolumn{3}{|c|}{$\chi$} \\
        \cline{2-4}
	& $2,048$ & $4,096$ & $8,192$ \\
	\hline
    	$11$ & $2.39\times10^{-5}$ & $5.10\times10^{-6}$ & $6.41\times10^{-7}$ \\
	\hline
    	$13$ & $1.87\times10^{-5}$ & $4.01\times10^{-6}$ & $5.08\times10^{-7}$ \\
	\hline
    	$15$ & $1.48\times10^{-5}$ & $3.16\times10^{-6}$ & $4.00\times10^{-7}$ \\
	\hline
    	$17$ & $1.21\times10^{-5}$ & $2.57\times10^{-6}$ & $3.26\times10^{-7}$ \\
	\hline
    	$19$ & $1.02\times10^{-5}$ & $2.17\times10^{-6}$ & $2.74\times10^{-7}$ \\
	\hline
    	$21$ & $8.77\times10^{-6}$ & $1.87\times10^{-6}$ & $2.36\times10^{-7}$ \\
	\hline
    	$23$ & $7.84\times10^{-6}$ & $1.64\times10^{-6}$ & $2.08\times10^{-7}$ \\
	\hline
    \end{tabular}
    \caption{The truncation erros in $2$D DMRG.}
    \label{errors}
\end{table}

\begin{table}[h]
    \centering
    \begin{tabular}{|c|c|c|c|}
	\hline
    	\multicolumn{4}{|c|}{$J_z=0.6$} \\
	\hline
        \multirow{2}*{$L_x$} & \multicolumn{3}{|c|}{$\chi$} \\
        \cline{2-4}
	& $2,048$ & $4,096$ & $8,192$ \\
	\hline
    	$11$ & $-52.31128073$ & $-52.32675360$ & $-52.33120591$ \\
	\hline
    	$13$ & $-59.55409964$ & $-59.57201159$ & $-59.57733794$ \\
	\hline
    	$15$ & $-66.76720764$ & $-66.78673057$ & $-66.79257242$ \\
	\hline
    	$17$ & $-73.97240347$ & $-73.99274493$ & $-73.99891747$ \\
	\hline
    	$19$ & $-81.17485452$ & $-81.19577544$ & $-81.20208757$ \\
	\hline
    	$21$ & $-88.37617798$ & $-88.39740626$ & $-88.40388350$ \\
	\hline
    	$23$ & $-95.57700851$ & $-95.59849368$ & $-95.60499496$ \\
	\hline
    \end{tabular}\\
    \begin{tabular}{|c|c|c|c|}
    	\hline
    	\multicolumn{4}{|c|}{$J_z=1$} \\
	\hline
        \multirow{2}*{$L_x$} & \multicolumn{3}{|c|}{$\chi$} \\
        \cline{2-4}
	& $2,048$ & $4,096$ & $8,192$ \\
	\hline
    	$11$ & $-76.79074819$ & $-76.79444193$ & $-76.79506575$ \\
	\hline
    	$13$ & $-88.81843623$ & $-88.82257977$ & $-88.82329538$ \\
	\hline
    	$15$ & $-100.8289033$ & $-100.8332598$ & $-100.8340180$ \\
	\hline
    	$17$ & $-112.8338499$ & $-112.8383246$ & $-112.8391054$ \\
	\hline
    	$19$ & $-124.8365616$ & $-124.8411020$ & $-124.8418960$ \\
	\hline
    	$21$ & $-136.8381993$ & $-136.8427837$ & $-136.8435863$ \\
	\hline
    	$23$ & $-148.8392695$ & $-148.8438775$ & $-148.8446859$ \\
	\hline
    \end{tabular}
    \caption{The groundstate energy calculated by $2$D DMRG.}
    \label{energy}
\end{table}

\section{A larger cylinder with $L_y=8$}\label{AppA}
\begin{figure}[h]
\centering
\includegraphics[width=0.9\linewidth]{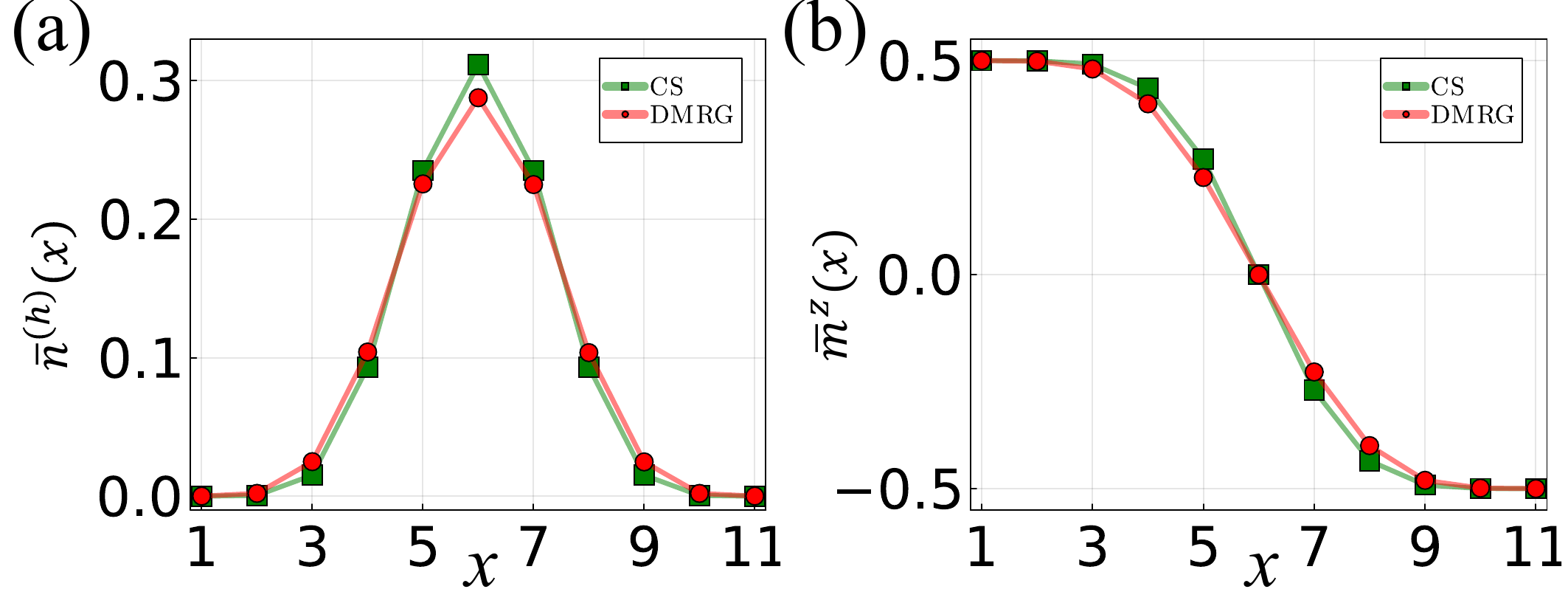}
\caption{The distribution of (a,b) the local hole density $n^{(\text{h})} (x,\ y)$ and (c,d) the local magnetic moment $m^z (x,\ y)$ in a $11\times8$ cylinder at $J_z=1$.
Their values are written in circles denoting the lattice sites in all rows.
A comparison is made between (a,c) the ED results of the effective model~\eqref{effHam} and (b,d) the DMRG results of the $t$-$J_z$ model~\eqref{eq1}.
 }\label{figs1}
\end{figure}

\begin{figure}[h]
\centering
\includegraphics[width=0.9\linewidth]{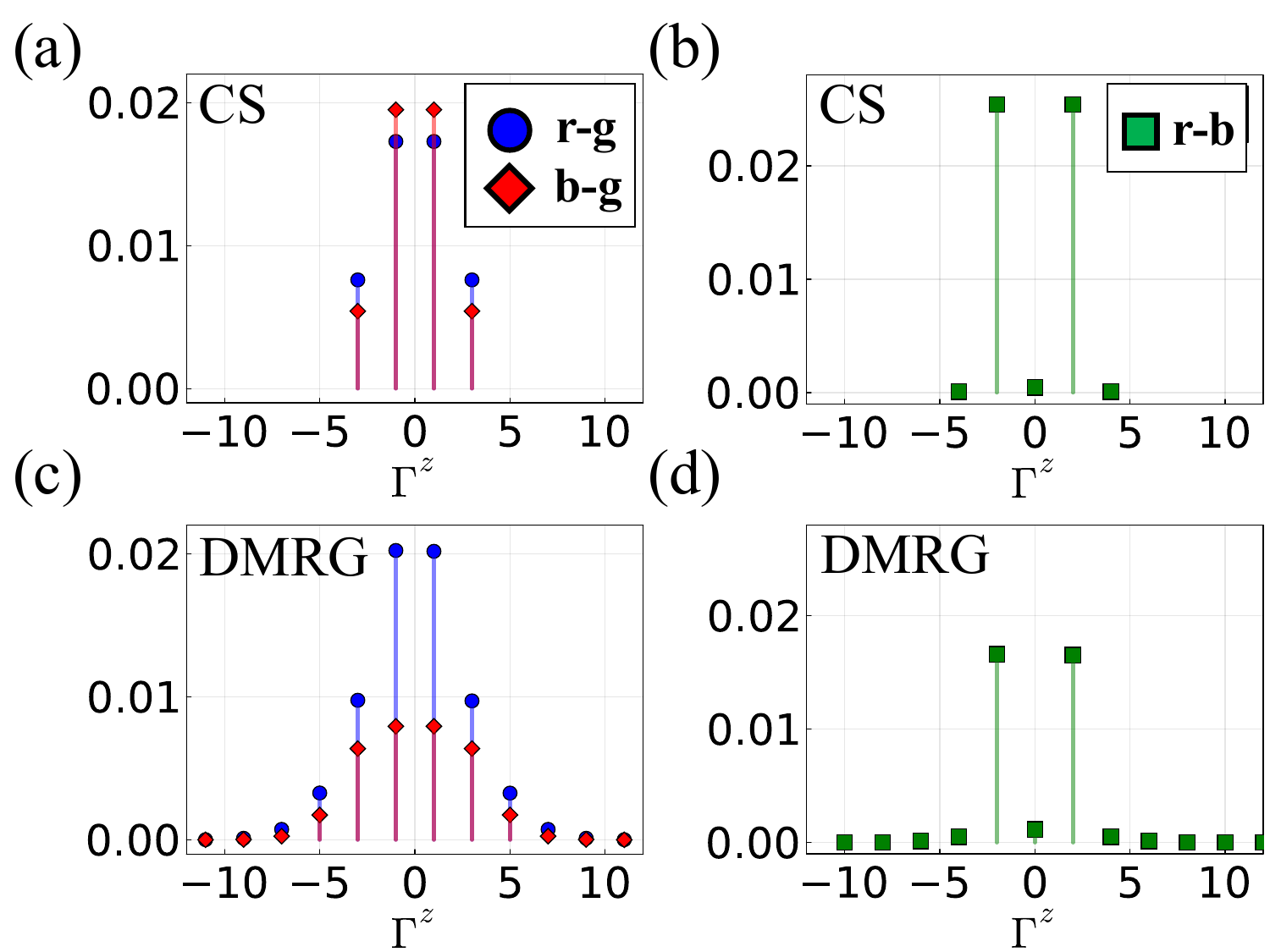}
\caption{
The spectrum weight $\Lambda \left(c, c', \Gamma^z \right)$ as a function of the spin field $\Gamma^z$ in a $11 \times 8$ cylinder at $J_z=1$.
Both the channels (a,b) \textbf{r}-\textbf{g} and \textbf{b}-\textbf{g}, as well as (c,d) \textbf{r}-\textbf{b}, are investigated.
A comparison is made between (a,c) the ED results of the effective model~\eqref{effHam}, and (b,d) the DMRG results of the $t$-$J_z$ model~\eqref{eq1}.
 }\label{figs2}
\end{figure}

To show the size effect caused by the finite circumference $L_y$, we consider a larger cylinder with $L_y=8$.
After truncating the effective spin field to $|\Gamma^z|\le6$, the resulting Hilbert space dimension is approximately $1,522,307,837 \approx 1.4 \text{G}$, which is notably large.
Due to the limitations of computer memory, we restrict $\vert \Gamma^z_{\bar{y}} \vert \le 4$ in the effective model~\eqref{effHam}.
As shown in Fig.~\ref{figs1}, both the distribution of the average hole density and magnetic moment demonstrate that the ED results from the effective model still match well with the DMRG results from the $t$-$J_z$ model, when the bond dimension $\chi=12,288$ is adopted.
Moreover, as shown in Fig.~\ref{figs2}, the spectra of the spin fields for $L_y=8$ also display the same features as the one for $L_y=6$.

\newpage

\bibliography{ref}

\end{document}